\documentclass[acmsmall,screen]{acmart}
\AtBeginDocument{%
  }

\usepackage{multirow}
\usepackage{float}
\usepackage{threeparttable}
\usepackage{tabularx}
\usepackage{colortbl}
\usepackage[most]{tcolorbox}
\newcommand{\inc}[1]{\cellcolor{green!15}{#1}}
\newcommand{\dec}[1]{\cellcolor{red!15}{#1}}
\newcommand{\same}[1]{\cellcolor{gray!15}{#1}}
\newcommand{\minv}[1]{\underline{#1}}
\newcommand{\maxv}[1]{\textbf{#1}}

\setcopyright{acmlicensed}
\copyrightyear{2026}
\acmYear{2026}
\acmDOI{XXXXXXX.XXXXXXX}

\acmJournal{JACM}
\acmVolume{37}
\acmNumber{4}
\acmArticle{111}
\acmMonth{8}

\begin{document}

\title{More than a Judge: An Empirical Study of Agent-Human Interaction in Crowdsourced Testing Assessment}

\author{Yue Wang}
\affiliation{%
  \institution{State Key Laboratory for Novel Software Technology, Nanjing University}
  \city{Nanjing}
  \country{China}
}
\email{yue_wang@smail.nju.edu.cn}

\author{Yuan Zhao}
\authornote{Corresponding author.}
\affiliation{%
  \institution{Laboratory of Data Intelligence and Interdisciplinary Innovation, Nanjing University}
  \city{Nanjing}
  \country{China}
}
\email{zhaoyuan@nju.edu.cn}

\author{Shengcheng Yu}
\affiliation{%
  \institution{Technical University of Munich}
  \city{Munich}
  \country{Germany}
}
\email{shengcheng.yu@tum.de}

\author{Zhenyu Chen}
\affiliation{%
  \institution{State Key Laboratory for Novel Software Technology, Nanjing University}
  \city{Nanjing}
  \country{China}
}
\email{zychen@nju.edu.cn}

\author{Qing Gu}
\affiliation{%
  \institution{State Key Laboratory for Novel Software Technology, Nanjing University}
  \city{Nanjing}
  \country{China}
}
\email{guq@nju.edu.cn}
\renewcommand{\shortauthors}{Wang et al.}

\begin{abstract}
Agentic AI is increasingly being integrated into software engineering workflows. In crowdsourced testing, however, the large volume and uneven quality of submitted reports still create a substantial review burden for developers. In prior work, we developed and validated a multi-agent assessment backbone based on the LLM-as-a-Judge paradigm. That backbone assesses reports along three dimensions---textuality, adequacy, and competitiveness---and was shown to align well with human consensus while substantially reducing assessment effort. Yet reliable automated judging does not by itself show whether agent outputs can improve human work when embedded into workflow. This paper studies that missing question in the context of crowdsourced testing. We investigate whether assessment-derived, actionable feedback can improve how testers revise reports, perform on later tasks, and transfer reporting practices across applications. To do so, we conducted a controlled four-stage human-subject study with 20 testers across three real-world applications. The results show that agent-generated feedback supports immediate improvements in revised reports, better first submissions on a new task after prior feedback exposure, and evidence of partial but meaningful transfer to a later application. A post-task questionnaire completed by 17 participants complements these artifact-based findings by suggesting that the feedback was generally understandable, acted upon in revision, and carried into later tasks, while also revealing remaining friction in specificity and execution. Overall, the study provides empirical evidence that, in the studied crowdsourced testing setting, assessment agents can serve not only as post-hoc judges but also as workflow-integrated feedback providers that support upstream report-quality improvement.
\end{abstract}

\ccsdesc[500]{Software and its engineering~Software testing and debugging}

\keywords{Crowdsourced Testing, Agent-Human Interaction, LLM-as-a-Judge, Large Language Models}

\received{November 2025}
\received[revised]{March 2026}
\received[accepted]{June 2026}

\maketitle

\section{Introduction}

Agentic AI is profoundly reshaping the landscape of software engineering (SE). The autonomous deployment of tools with Large Language Model (LLM) backends gives agents tremendous potential to bring greater automation to core SE tasks, including testing and analysis\cite{hosseini2025role}. Crowdsourced testing, in particular, is a critical SE task that stands to benefit from this transformation. Crowdsourced testing leverages a diverse, large-scale workforce to enhance test coverage and the discovery rate of defects compared to traditional methods~\cite{mao2017survey,gao2019successes}. However, this benefit is coupled with a significant challenge: crowdsourced testing platforms receive a massive volume of test reports from crowdworkers of varying backgrounds~\cite{feng2016multi,hao2019ctras,li2022classifying}. This influx of reports, often of inconsistent quality, creates a severe bottleneck for developers, making manual review time-consuming, labor-intensive, and prone to subjectivity~\cite{feng2015test,yu2021prioritize,aversano2016bug,zimmermann2010makes,bhattacharya2013empirical}. At the same time, as agents become more capable of participating in SE workflows, an additional question emerges. Beyond automating judgment, it becomes important to understand whether agent outputs can be trusted, interpreted, and integrated into iterative human work. In crowdsourced testing, where quality improvement often depends on feedback, revision, and coordination across contributors, this means that the role of agent-generated feedback should be studied not only as an assessment mechanism, but also as a workflow and agent-human interaction mechanism.

To alleviate the report assessment bottleneck itself, automated quality assessment methods have emerged~\cite{chen2018automated,chen2020systemic,zhang2022automated}. In our prior work, we proposed a multi-dimensional assessment method for crowdsourced testing reports based on LLMs~\cite{ase2025multi}. This method is operationalized as a multi-agent framework where specialized agents, acting as LLM-as-a-Judge~\cite{zheng2024judging}, assess reports from three complementary dimensions: Textuality, which examines the fine-grained textual quality of an individual report; Adequacy, which measures functional coverage against test requirement documents; and Competitiveness, which assesses the novelty and value of discovered defects. That prior study focused on validating the agents as reliable and efficient judges. It established a necessary foundation for practice by showing that the agents can produce assessments aligned with human consensus while substantially reducing assessment effort, thereby addressing a trust-related prerequisite for deploying such agents in real workflows.

However, validating automated judging capability does not by itself show that agentic AI improves human performance. In software work, feedback is valuable only when people can understand it, decide to act on it, and carry what they learn into subsequent tasks. This gap is especially important for agentic AI in software: once agents move from isolated benchmarking settings into existing workflows, the central question shifts from what the agents can score to how humans respond to agent assessments. For crowdsourced testing, the missing scientific question is therefore not whether the assessment agents can score reports again, but whether agent-generated feedback changes testers' revision behavior, performance on later tasks, and the transfer of reporting skills.

Compared with our prior work, this paper does not aim to further validate the assessment agents themselves. Prior work established that the agents can assess crowdsourced testing reports with high agreement with human consensus and with substantial efficiency gains, providing the basis for trusting their assessments in practice. In contrast, this paper investigates a different research question: whether these already-validated agents can improve human testers when their assessments are delivered as actionable feedback in a realistic workflow. Accordingly, the main extension of this paper lies not in re-evaluating the same scoring framework, but in a new human-subject study with a different unit of analysis, different outcomes, and different claims. More broadly, this paper shifts the focus from autonomous assessment capability to agent-human interaction, workflow integration, and the role of feedback in shaping individual performance and team-facing work practices.

We therefore study agent-generated feedback as a workflow intervention. We designed and executed a controlled human-subject study involving 20 testers, divided into two groups (A and B), who performed testing tasks across four stages using three real-world applications. Both groups first completed the same baseline task on APP1. The intervention was then staggered across the two groups: Group A received agent-generated feedback on its APP1 reports and revised those reports before moving to APP2, whereas Group B first completed APP2 without feedback and only afterward received agent-generated feedback on its APP2 reports for revision. After each group had experienced one feedback cycle, both groups completed a final task on APP3. This design lets us examine immediate revision effects, new-task performance after prior feedback exposure, and longer-term skill transfer in a realistic SE process. We address the following three research questions (RQs):

\begin{itemize}
  \item \textbf{Effect of Agent-Generated Feedback on Report Revision}. In RQ1, we evaluate whether agent-generated feedback helps testers improve the quality of their existing reports within the same task. Results: Testers who received feedback substantially improved the Textuality and Adequacy scores of their revised reports, indicating immediate benefits from the feedback loop.
  \item \textbf{Impact on Performance in a New Task}. In RQ2, we investigate whether prior interaction with agent feedback influences a tester's initial performance on a new application. Results: The group with prior feedback exposure produced higher-quality initial reports on the new task, suggesting that part of what they learned from the feedback carried over beyond the original revision setting.
  \item \textbf{Transfer of Skills Across Tasks}. In RQ3, we examine whether repeated exposure to agent-generated feedback is associated with longer-term skill transfer on a later, distinct application. Results: We observe evidence of positive transfer, with later-task performance improving relative to the initial baseline, suggesting that the feedback can support more durable capability development in the studied setting.
\end{itemize}

This paper provides empirical evidence, in the studied setting, on how agent-generated feedback affects human performance in crowdsourced testing workflows. The main contributions of this paper are:

\begin{itemize}
  \item \textbf{A workflow-oriented reframing of assessment agents.} We study how previously validated assessment agents operate as feedback providers in a human-centered crowdsourced testing workflow, rather than re-validating them as judges. This reframing positions the work around agent-human interaction, workflow integration, and the practical conditions under which trusted agent assessments can improve human work.
  \item \textbf{A new controlled human-subject methodology.} We conduct a staggered four-stage study with 20 testers, two groups, and three real-world applications, enabling a structured analysis of immediate report revision, performance on an unseen task, and subsequent skill transfer after exposure to agent-generated feedback.
  \item \textbf{New evidence on human impact in feedback-oriented crowdsourced testing.} We provide evidence that integrating assessment agents into the feedback loop can improve revised report quality and support better performance on later tasks, suggesting a practical path by which agent-generated feedback can contribute to workflow-level improvement and capability development in crowdsourced testing.
\end{itemize}

\section{Background and Related Work}
\subsection{Crowdsourced Test Reports}
\label{2.1}
\begin{figure}[h]
  \centering
  \includegraphics[width=\linewidth]{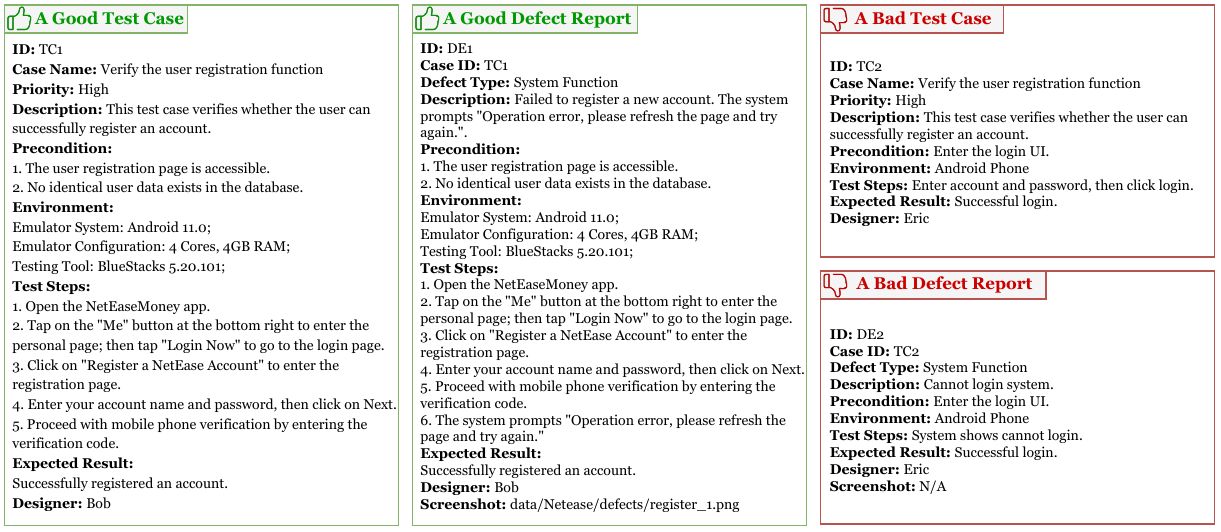}
  \caption{Example of Crowdsourced Test Reports: The Good and The Bad.}
  \label{fig:report-example}
\end{figure}

With the increasing scale and complexity of software, crowdsourced testing has emerged as an effective approach to enhance test coverage and defect discovery~\cite{gao2019successes}. It leverages the collective intelligence of distributed testers from diverse backgrounds. The critical deliverable in the crowdsourced testing process is the crowdsourced testing report~\cite{yu2021prioritize}. These reports are the detailed records of test activities and are essential for developers to understand and fix defects~\cite{feng2015test}. Each report typically consists of two main components: test cases designed by the tester and defects discovered during test execution. While formats may vary across different crowdsourcing platforms, core fields remain consistent, such as testing environment specifications, step-by-step test procedures, and actual results~\cite{feng2016multi}. However, the quality of these reports varies dramatically. As illustrated in Figure~\ref{fig:report-example}, a "Good Test Case" (TC1) contains a descriptive name, well-defined preconditions, logically structured steps, and clear expected results. In contrast, a "Bad Test Case" (TC2) is characterized by ambiguous descriptions and insufficient specifications, which hampers reproducibility. Similarly, a "Good Defect Report" (DE1) is clear, logical, and reproducible, while a "Bad Defect Report" (DE2) has ambiguous descriptions that impede a developer's ability to resolve the issue. The large volume and inconsistent quality of these reports necessitate automated assessment methods to help developers filter and review them efficiently.
\label{sec:2.1}

\subsection{Processing and Assessing Crowdsourced Test Reports}
\label{2.2}
The challenges posed by the volume and variable quality of crowdsourced testing reports have led to a significant body of research focused on automated processing. These efforts aim to assist developers in managing the influx of reports and improving review efficiency. Key research directions include Test Report Prioritization, which aims to rank reports so that developers can focus on the most valuable ones first, often using features from screenshots and text~\cite{feng2015test,feng2016multi,yu2021prioritize}; Report Clustering and Summarization, which groups similar reports to manage redundancy, such as using features from screenshots and text for semi-supervised clustering~\cite{yu2024semi}; Report Deduplication, which focuses on identifying duplicate reports using information retrieval models to measure similarity~\cite{alipour2013contextual,hindle2016contextual,nguyen2012duplicate,sun2011towards,sun2010discriminative}; and Quality Assessment, where studies have modeled test reports based on predefined quality metrics to help developers focus on high-quality submissions~\cite{chen2018automated,chen2020systemic}. While these automated methods improve efficiency, they often lack comprehensive assessment dimensions and, crucially, do not provide actionable feedback mechanisms for the testers themselves, which restricts these methods to passively filtering information, ultimately failing to address the problem of low-quality test reports at its source.

\subsection{Agents as Assessors (LLM-as-a-Judge)}
\label{2.3}
The emergence of the LLM-as-a-Judge (LLM-J) paradigm~\cite{zheng2024judging,chan2023chateval,liu2023g} offers a powerful new approach for automated assessment. This paradigm leverages the sophisticated natural language understanding capabilities of LLMs to serve as automated evaluators for both objective and subjective tasks. Unlike traditional reward models that only output a scalar score, LLM-J evaluators are considered more robust and interpretable because they can generate detailed rationales and justifications for their assessments. Recent studies demonstrate the effectiveness of LLM-J across diverse applications. In software engineering, this includes code generation evaluation~\cite{ahmed2024can,wang2025can}, code summarization assessment~\cite{wu2024can}, and automated grading of short answers~\cite{chang2024automatic,henkel2024can,senanayake2024rubric,fagbohun2024beyond,jury2024evaluating}. These studies consistently show that LLM-J assessments can achieve high consistency with human evaluators while significantly reducing time and cost. Our prior work (the basis of this extension) applied this paradigm to the assessment of crowdsourced testing reports, demonstrating its high consistency with human expert consensus.

\subsection{Agent-Human Interaction in Software Engineering}

The previous subsections describe the problem domain (\ref{2.1}), traditional report-processing solutions (\ref{2.2}), and the assessment foundation used in this paper (\ref{2.3}). However, greater automation does not automatically imply greater human productivity. A growing line of work therefore studies how agents are integrated into software-engineering workflows and how humans respond to agent outputs in practice~\cite{feng2024large}. Examples include agent-assisted development\cite{manish2024autonomous}, where systems such as GitHub Copilot\cite{GitHubCopilot} provide code-generation and completion support during implementation; agent-assisted code review, where automated tools and bots provide review-time feedback in CI/CD workflows~\cite{tang2024codeagent,rasheed2024ai}; and agent-assisted education and training, where automated feedback helps users identify and correct errors during learning activities\cite{wang2025impact,guo2024using}. Together, these studies suggest that the practical value of agents depends not only on autonomous capability, but also on how agent outputs are interpreted, trusted, and acted upon in workflow.

In the context of LLM-as-a-Judge, however, most research has concentrated on validating the quality of the judgments themselves. Much less is known about what happens when assessment outputs are delivered back to humans as workflow-integrated feedback. In crowdsourced testing, this gap is especially important: a detailed, multi-dimensional assessment may be valuable not only for downstream review support, but also for helping testers revise reports and carry improved practices into later tasks. The present study addresses this feedback-oriented question. Rather than asking again whether the agents can judge reports well, we examine whether assessment-derived feedback can function as a practical intervention in the crowdsourced testing workflow and thereby influence revision behavior, later-task performance, and partial skill transfer.

Viewed together, these strands of prior work motivate a methodological shift in the present paper. Once assessment outputs are embedded into workflow, the central question is no longer only whether agents can judge reports well, but also whether their feedback can produce observable improvements in human work. The next section therefore summarizes the multi-dimensional assessment backbone and explains how it is instantiated here as a feedback-oriented workflow intervention.

\section{The Multi-dimensional Assessment Agents in a Feedback-Oriented Workflow}
\label{sec3}
\begin{figure}[h]
  \centering
  \includegraphics[width=\linewidth]{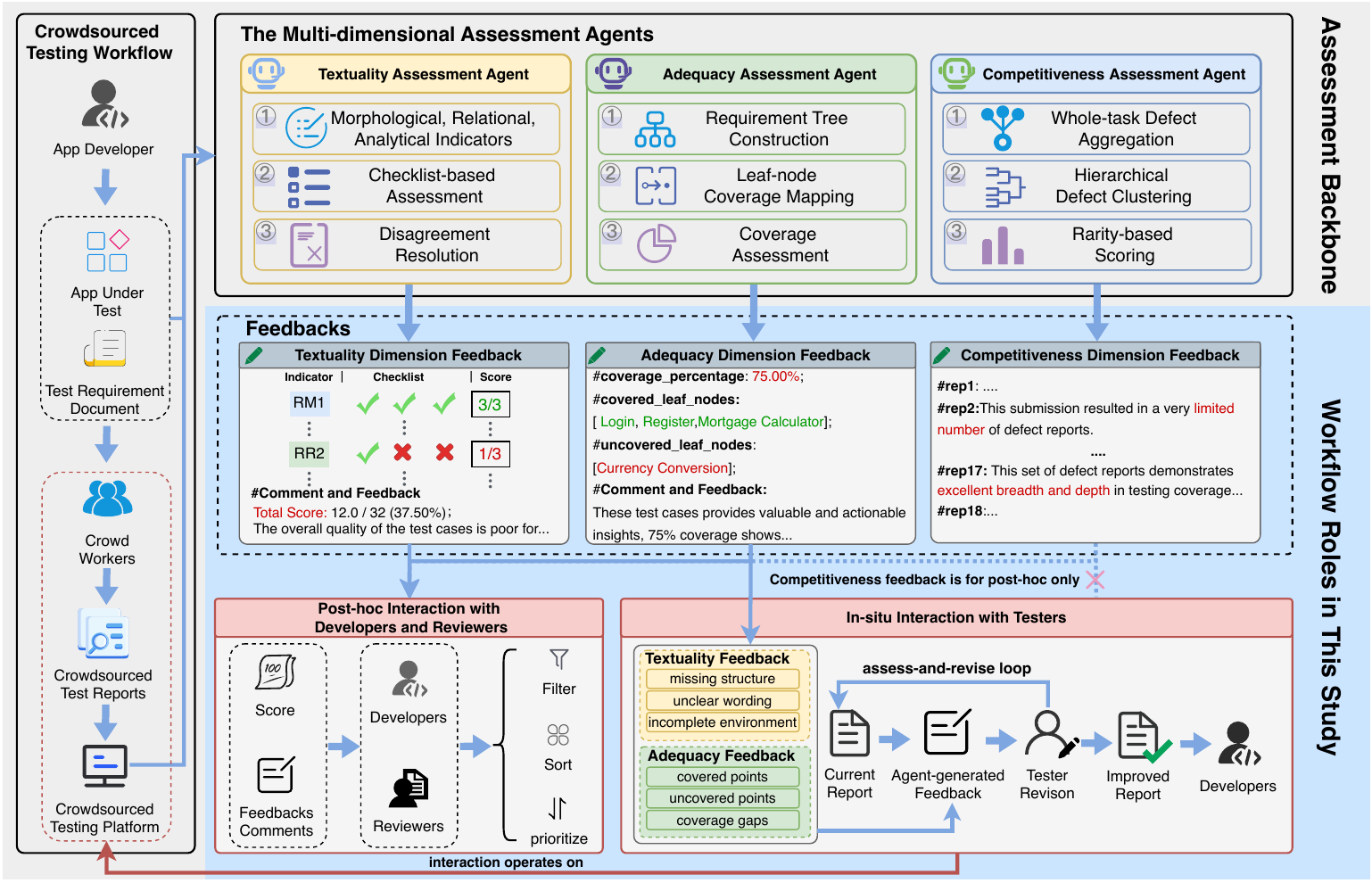}
  \caption{The Multi-dimensional Assessment Agents and their workflow roles in this study.}
  \label{fig:framework}
\end{figure}

\subsection{Overview}
Our human-subject study (described in Section \ref{sec4}) builds on the multi-agent assessment framework that we previously developed and validated~\cite{ase2025multi}. The assessment agents themselves are not newly proposed in this article. Instead, they are reused as the assessment backbone for report scoring and feedback generation in the present study. Accordingly, this section does not attempt to re-establish the validity of the agents as judges. Rather, it summarizes the essential mechanisms needed to understand how trusted assessment outputs are translated into human-facing feedback and embedded into an agent-human interaction loop.

In prior work, we benchmarked this backbone against six experienced software-testing experts across three crowdsourced testing tasks. Although individual human ratings showed only moderate consistency, the backbone aligned strongly with averaged expert consensus while also delivering substantial efficiency gains. In particular, prior work reported QWK values of 0.70--0.82 for textuality and 0.89--0.97 for adequacy against averaged human ratings, together with 5--90$\times$ assessment-efficiency improvements over manual review~\cite{ase2025multi}. These results established a sufficiently trustworthy measurement foundation for the present paper, whose focus is not further score validation but workflow-level human impact.

As illustrated in Figure~\ref{fig:framework}, Section~\ref{sec3} is organized into two methodological layers. The first layer is the multi-dimensional assessment backbone, which comprises the Textuality, Adequacy, and Competitiveness agents. The second layer is the feedback-oriented workflow studied in this paper, where agent outputs are operationalized as actionable intervention for human testers. In this setting, Textuality and Adequacy support the in-situ intervention because both can be computed from an individual tester's current report and converted into revision guidance. Competitiveness is retained as part of the overall framework and remains useful for post-hoc workflow support, but it is not used as the intervention because it depends on cross-tester comparison after all reports have been collected. This distinction between assessment as measurement and assessment as intervention is central to our workflow-oriented methodology for agentic AI in software engineering.

\subsection{The Multi-dimensional Assessment Agents}
The reused assessment backbone comprises three specialized agents that assess crowdsourced testing reports from the complementary dimensions of Textuality, Adequacy, and Competitiveness. Together, they provide a structured basis for assessing report quality from the perspectives of writing quality, requirement coverage, and comparative defect value. In this subsection, we summarize only the key mechanisms needed for understanding the present study; the feedback-oriented workflow role of these outputs is described in Section~\ref{sec3.4}.

\subsubsection{Textuality Agent}
The Textuality Agent assesses the fine-grained quality of each test case and defect report. It operationalizes five desirable properties of a good report (atomicity, completeness, conciseness, understandability, and reproducibility) \cite{chen2018automated} through the quantitative indicators listed in Table~\ref{tab:metrics_optimized_final_centered_indicator}. These indicators are grouped into Morphological, Relational, and Analytical categories. In the present study, this dimension provides a stable measurement of report quality while also serving as a source of revision-oriented feedback.

    \begin{table}[h]
      \centering
      \caption{The Relationship Between Desirable Properties and Measurable Indicators}
      \label{tab:metrics_optimized_final_centered_indicator}
      \footnotesize 
      \scalebox{0.88}{
      \begin{threeparttable}
      \begin{tabular}{ c | c | >{\centering\arraybackslash}m{5em} | >{\centering\arraybackslash}m{5em} | >{\centering\arraybackslash}m{6em} | >{\centering\arraybackslash}m{8em} | >{\centering\arraybackslash}m{7em} }
      \hline
      \multirow{2}{*}{\textbf{Category}} & \multirow{2}{*}{\textbf{Indicator}} & \multicolumn{5}{c}{\textbf{Desirable Property}} \\
      \cline{3-7}
       &  & \textbf{Atomicity} & \textbf{Conciseness} & \textbf{Completeness} & \textbf{Understandability} & \textbf{Reproducibility} \\
      \hline
      \multirow{3}{*}{Morphological} & Text Length & $\bullet$ & $\times$ &  & $\bullet$ &  \\
      \cline{2-7}
       & Readability &  &  &  & $\times$ &  \\
      \cline{2-7}
       & Punctuation &  & $\times$ &  & $\bullet$ &  \\
      \hline
      \multirow{6}{*}{Relational} & Itemization & & $\bullet$ & $\bullet$ & $\times$  &  $\times$\\
      \cline{2-7}
       & Environment &  &  & $\times$ &  & $\times$ \\
      \cline{2-7}
       & Preconditions &  &  & $\times$ &  & $\times$ \\
      \cline{2-7}
       & Expected Results &  &  & $\times$ &  & $\times$ \\
      \cline{2-7}
       & Additional Information &  &  & $\times$ &  &    \\
      \cline{2-7}
       & Screenshot &  &  & $\times$ & $\times$ & $\times$ \\
      \hline
      \multirow{3}{*}{Analytical} & Interface Elements &  & $\bullet$ &  & $\times$ & $\times$ \\
      \cline{2-7}
       & User Behavior &  &  & $\bullet$ & $\times$ & $\times$  \\
      \cline{2-7}
       & Defect Response & &  & $\bullet$ &  & $\times$ \\
      \hline
      \end{tabular}
      \begin{center}
          \begin{tablenotes}
          \centering
          \footnotesize
          \item '$\times$' represents a direct influence and '$\bullet$' denotes an indirect influence between the indicator and the desirable property
      \end{tablenotes}
      \end{center}
      \end{threeparttable}
      }
      \label{tab:metrics}
      \end{table}

To keep this dimension sufficiently stable for human-facing use, the agent preserves two mechanisms from the previously validated framework. First, to overcome the limitation of LLMs in providing accurate continuous scores \cite{liu2023g}, the agent's judgment is guided by a Checklist, following the design principles of RocketEval~\cite{wei2025rocketeval}. This checklist decomposes the assessment task into a series of binary checkpoints corresponding to each indicator, where the number of checkpoints matches the indicator's score value. Following the practice of Zhang et al.'s work~\cite{zhang2022automated}, we defined score upper limits for each indicator based on the relationship between desirable properties and quantitative indicators, as shown in Table~\ref{tab:rule_summary}. This design turns a continuous judgment problem into a set of interpretable pass/fail decisions and preserves the local evidence needed for later feedback generation.

\begin{table}[t]
  \caption{Quantitative Indicators Definitions} 
  \label{tab:rule_summary}
  \footnotesize
  \centering
  \scalebox{0.95}{
  \begin{tabular}{lllll}
  \toprule
  \textbf{Category} & \textbf{Indicators} & \textbf{Rule} & \textbf{Rule Content} & \textbf{Score} \\
  \midrule
  \multirow{3}{*}{Morphological}& Text Length & RM1 & Text length is within the preset range. & 3 \\
  & Readability & RM2 & The description is concise, fluent, and easy to understand. & 2 \\
  & Punctuation & RM3 & Punctuation is used correctly. & 3 \\
  \midrule
  \multirow{6}{*}{Relational}& Itemization & RR1 & Operational steps are listed with numbered annotations. & 5 \\
  & Environment & RR2 & Environmental information is present and detailed. & 3 \\
  & Preconditions & RR3 & Preconditions are described and complete. & 2 \\
  & Expected Results & RR4 & Expected results are filled in standardly. & 2 \\
  & Additional Information & RR5 & All other fields except for the above information are filled in. & 2 \\
  & Screenshot & RR6(*) & For defects, there should be screenshots for explanation. & 3 \\
  \midrule
  \multirow{3}{*}{Analytical} & User Interface & RA1 & There is a clear enough description of the interface elements. & 5 \\
  & User Behavior & RA2 & There is a description of interactive behavior. & 5 \\
  & Defect Feedback & RA3(*) & For defects, defect feedback should be included. & 3 \\
  \bottomrule
  \end{tabular}
  }
  \begin{center}
      \begin{tablenotes}
      \footnotesize
      \item Rules with an asterisk (*) are for defects, due to differences in scoring criteria from test cases.
  \end{tablenotes}
  \end{center}
  \end{table}

Second, the agent implements a Dual LLM Assessment and Disagreement Resolution mechanism. Two independent LLMs (DeepSeek and Kimi) simultaneously assess the same report against the checklist. When disagreements arise, the agent itself (using gpt-4o as a base) performs a "thinking and judgment" process to analyze the conflict and produce a unified, final checklist result. The final textuality score is then calculated by summing all checkpoints marked as 'True', as shown in the following formula:

\begin{equation}
\label{eq:textuality_score}
TextualityScore = \sum_{i=1}^{N} \sum_{j=1}^{|C_i|} \mathbb{I}(c_{i,j} = \text{True})
\end{equation}

where $N$ represents the total number of indicators, $C_i$ denotes the set of checkpoints for indicator $i$, and $\mathbb{I}(c_{i,j} = \text{True})$ is an indicator function that equals 1 when checkpoint $j$ of indicator $i$ passes and 0 otherwise. This approach provides a cumulative score reflecting the total number of quality criteria satisfied by the report. In the present study, the score is used for measurement, while the underlying checklist findings later support report-level revision guidance.

\subsubsection{Adequacy Agent}
The Adequacy Agent assesses how comprehensively a report covers the specified functional requirements. Requirement documents are the grounding artifact for this dimension in crowdsourced testing~\cite{feng2015test}, and coverage remains a basic indicator of how thoroughly test cases examine software functionality~\cite{tran2025quality}. In the present study, this dimension is especially important because it makes requirement coverage traceable and exposes the functional gaps that can later be returned to testers as actionable feedback.

The agent's process involves three sequential steps. 
\textbf{Requirement Tree Construction} systematically transforms unstructured requirement documents into hierarchical trees where leaf nodes represent atomic functional points. This phase performs structural analysis to identify major functional modules and their relationships, decomposes compound requirements into atomic units, and then constructs a hierarchy whose leaf nodes can be directly mapped to test cases.

\textbf{Test Case Mapping} establishes connections between individual test cases and the atomic functional points in the requirement tree. For each submitted test case, the agent uses LLM-based semantic analysis to compare the test case content (e.g., name, description, steps) against the list of all functional leaf nodes from the requirement tree, thereby identifying coverage relationships that might be missed by keyword-based or rule-based approaches.

\textbf{Coverage Assessment} synthesizes the mapping results for all test cases within a single crowdsourced testing report. It calculates the total functional coverage percentage using the following formula:
\begin{equation}
\label{eq:adequacy_score}
\text{Coverage}(R) = \frac{|\bigcup_{T_i \in R} \text{CoveredLeafNodes}(T_i)|}{|\text{LeafNodes}|} \times 100\%
\end{equation}
where $R$ represents a crowdsourced testing report, $T_i$ denotes an individual test case, and $|\text{LeafNodes}|$ is the total number of leaf nodes in the requirement tree. Here, $\text{CoveredLeafNodes}(T_i)$ denotes the set of leaf nodes covered by test case $T_i$. This formulation provides a coverage-based score for analysis while preserving the mapping evidence that later supports feedback on which functional points have been covered and which leaf nodes remain unaddressed.

\subsubsection{Competitiveness Agent}
The Competitiveness Agent addresses the comparative nature of crowdsourced testing, where valuable outcomes include novel and rare defects \cite{mao2017survey}. It performs a task-level analysis across all submitted reports to assess the rarity and uniqueness of the defects discovered by each tester.

The process begins with \textbf{Defect Aggregation}, where the agent collects all individual defect reports submitted for the same testing task. Next, it performs \textbf{LLM-based Hierarchical Clustering}. The agent instructs an LLM to analyze all defects based on their root causes, observable symptoms, and functional impact, and to automatically organize similar defects into a fine-grained, multi-level clustering tree. This semantic clustering can group defects that have different surface-level descriptions but share the same underlying cause. 

Following clustering, the agent performs \textbf{Defect Scoring} using a rarity-based mechanism, as shown in the following formula:
\begin{equation}
\label{eq:competitiveness_score}
Score(d) = \frac{S_{max}}{|N_d|}
\end{equation}
where $S_{max}$ is the maximum possible score, and $|N_d|$ is the number of defect reports in the leaf node containing defect $d$. Defects that are clustered into leaf nodes containing fewer reports (i.e., those discovered by fewer testers) are considered rarer and are assigned significantly higher scores. 

Finally, the agent computes the \textbf{Report Score Aggregation} by summing the scores of all unique defect categories (leaf nodes) that a tester contributed to, ensuring that a tester receives the score for a specific defect category only once, even if they reported multiple instances of the same bug. We retain this dimension because it remains part of the full assessment backbone and still supports post-hoc review and prioritization. However, unlike Textuality and Adequacy, Competitiveness depends on the complete defect landscape across all testers and therefore is not used as the tester-facing intervention in this study.

\subsection{Feedback Generation and Agent-Human Interaction}
\label{sec3.4}
This paper focuses on the feedback-oriented instantiation of the assessment backbone. As illustrated in Figure~\ref{fig:framework}, the framework supports two complementary interaction paradigms, targeting different stakeholders at different stages of the testing workflow.

The first mode of interaction is \textbf{post-hoc, for developers and reviewers}. In this paradigm, the agents' assessments are utilized \textit{after} the test reports have been submitted. The multi-dimensional outputs support review, filtering, sorting, and prioritization, thereby helping developers manage the large volume of incoming reports. This interaction type uses assessment results to streamline downstream workflow management.

The second mode of interaction is \textbf{in-situ, for crowdsourced testers}, and it is the focus of our human-subject study. In this paradigm, the intervention is not the raw score alone but the actionable feedback derived from Textuality and Adequacy assessments and returned to testers during revision. For Textuality, the system converts checklist findings into report-level guidance about missing structure, ambiguous wording, incomplete preconditions, expected results, or other weak elements. For Adequacy, it converts requirement-tree mapping results into coverage guidance by showing which functional points have been exercised and which leaf nodes remain uncovered. What testers receive is therefore a human-facing feedback package that tells them what to rewrite, complete, or extend in the current report, rather than merely how they scored.

This distinction is important because the intervention is the feedback content, not the score alone. Textuality findings become revision-oriented guidance on report quality, while uncovered requirement leaf nodes become adequacy-oriented guidance on coverage gaps. Competitiveness is intentionally excluded from this in-situ package because it requires the complete cross-tester defect set and is therefore more appropriate for post-hoc comparison than local revision support.

This assess-and-revise loop is the methodological focus of the present article. Because the assessment backbone was already validated~\cite{ase2025multi}, the current study can concentrate on a different question: whether agent-generated feedback can be meaningfully integrated into a realistic software workflow and thereby influence human performance, revision behavior, and later skill transfer. Section~\ref{sec4} therefore evaluates this in-situ interaction as a human-subject, workflow-oriented intervention rather than as a further benchmark of agent scoring accuracy.

\section{Study Design}
\label{sec4}
This section presents the human-subject methodology used to evaluate the workflow-level impact of integrating the feedback-oriented assessment backbone into crowdsourced testing. In prior work, the emphasis was on the reliability and efficiency of the assessment backbone as an evaluation instrument~\cite{ase2025multi}. Here, the focus is different: we study whether actionable feedback derived from that backbone can improve tester performance during report production and revision, both within the same task and across later tasks.

\subsection{Study Overview and Research Questions}
To operationalize the in-situ tester-facing interaction mode described in Section~\ref{sec3}, we designed a controlled four-stage study with two tester groups and three applications. The design supports three complementary comparisons: immediate within-round revision after feedback, first-submission performance on a new task after prior feedback exposure, and longer-term transfer after both groups have experienced at least one feedback cycle. The study addresses the following research questions:

\begin{itemize}
  \item \textbf{RQ1 (Effect of Agent-Generated Feedback on Report Revision):} How effective is agent-generated feedback in helping testers improve the quality of their existing reports within the same task?
  \item \textbf{RQ2 (Impact of Prior Feedback Exposure on New-Task Performance):} How does prior exposure to agent-generated feedback affect a tester's initial performance on a new testing task?
  \item \textbf{RQ3 (Transfer of Reporting Practices Across Tasks):} To what extent do practices reinforced by agent-generated feedback transfer to subsequent, different testing tasks?
\end{itemize}

\subsection{Participants and Grouping}
We recruited 20 testers for this study from a major crowdsourced testing platform called MoocTest\cite{mooctest2025} in China. To better characterize the participant pool, 17 of the 20 participants completed a post-task questionnaire. The respondent subset is best characterized as a trained cohort of crowd testers with heterogeneous backgrounds, rather than as either complete novices or professional software testers. All respondents reported formal software-testing coursework, and most also reported systematic training in testcase design and defect reporting. At the same time, meaningful variation was observed in prior crowdsourced-testing duration, familiarity with the three application domains used in the study, programming and software-project experience, and routine AI-tool usage. Table~\ref{tab:participant_profile} summarizes these background characteristics for the respondent subset.

Based on historical platform performance and scores, we divided the full 20-person study sample into two groups of 10 testers each: Group A and Group B. This grouping strategy was intended to keep baseline skill levels as comparable as possible before the intervention.

\begin{table}[t]
  \caption{Background characteristics of the questionnaire respondent subset ($n=17$)}
  \label{tab:participant_profile}
  \small
  \centering
  \setlength{\tabcolsep}{3pt}
  \renewcommand{\arraystretch}{1.08}
  \begin{tabularx}{\linewidth}{@{}>{\raggedright\arraybackslash}p{0.26\linewidth} >{\raggedright\arraybackslash}X@{}}
    \toprule
    \textbf{Aspect} & \textbf{Summary} \\
    \midrule
    Questionnaire coverage & 17/20 participants responded (85.0\%); 8/10 from Group~A and 9/10 from Group~B. \\[2pt]
    Formal testing preparation & All 17 reported formal software-testing coursework; testcase-design training: 17/17; defect-report training: 16/17 (one with limited classroom exposure only). \\[2pt]
    CT experience duration & $<$3 months: 5/17 (29.4\%); 3--6 months: 10/17 (58.8\%); 6--12 months: 1/17 (5.9\%); $>$1 year: 1/17 (5.9\%). \\[2pt]
    Req.-oriented testing habits & Familiarity mean = 4.00/5; 15/17 (88.2\%) rated 4--5; 13/17 (76.5\%) mainly designed from requirement documents; 15/17 (88.2\%) regularly decomposed requirements into function points. \\[2pt]
    Domain familiarity & Management/system: 4--5 by 14/17 (82.4\%); e-commerce: 8/17 (47.1\%); utility/to-do: 9/17 (52.9\%). \\[2pt]
    Technical background & Programming experience: 16/17 (94.1\%); prior project participation: 5/17 (29.4\%); web/Android experience mixed between usage-only and usage-plus-development profiles. \\[2pt]
    Routine AI-tool usage & Occasional use: 11/17 (64.7\%); frequent use: 6/17 (35.3\%). \\
    \bottomrule
  \end{tabularx}
\end{table}

In this subsection, we use the questionnaire only to clarify the background characteristics of the respondent subset and thereby better contextualize the participant pool. More specific questionnaire findings related to feedback understanding, trust, adoption, transfer, and possible external-AI confounds are reported later in Section~\ref{sec:questionnaire-findings} as complementary evidence for interpreting the main artifact-based results.

\subsection{Study Materials and Feedback Intervention}
The study used three real-world applications representing distinct crowdsourced-testing contexts: APP1, a management system; APP2, an online shopping platform; and APP3, a utility app. Each application was accompanied by a task-specific PDF requirement document distributed as part of the testing task. These documents described the relevant functional scope, user flows, and test targets for the application under test. In the present study, they grounded the Adequacy dimension by enabling requirement-tree construction and leaf-node coverage mapping, and they also provided the reference material used in the feedback intervention.

The intervention in this study was a human-facing feedback package derived from the assessment backbone rather than the raw scores alone. For Textuality, the package translated checklist findings into revision guidance on issues such as missing structure, ambiguous wording, incomplete preconditions, expected results, or other weak report elements. For Adequacy, it translated requirement-tree mapping results into coverage guidance by showing which functional points had already been exercised and which leaf nodes remained uncovered. Testers therefore received actionable instructions about what to rewrite, complete, or extend in the current report. Following the in-situ interaction logic described in Section~\ref{sec3}, Competitiveness was intentionally excluded from the intervention because it depends on the complete cross-tester defect set and is therefore inherently post-hoc.

\subsection{Four-Stage Experimental Procedure}
\begin{figure}[h]
  \centering
  \includegraphics[width=\linewidth]{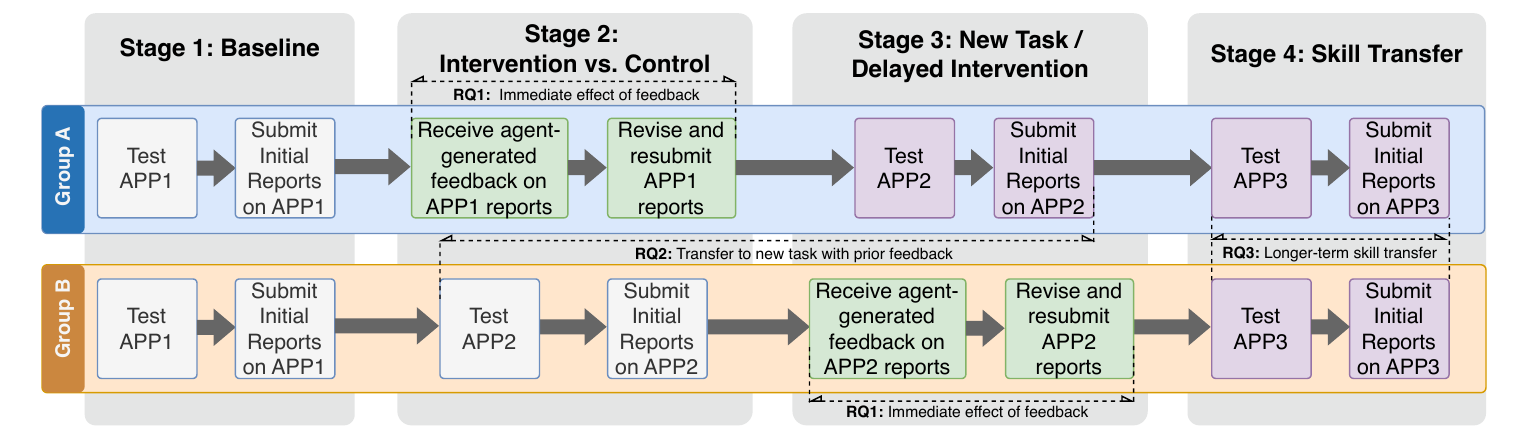}
  \caption{Four-stage human-subject workflow for evaluating in-situ agent-generated feedback in crowdsourced testing. The design supports three complementary comparisons: within-round revision after feedback (RQ1), first-submission quality on a new task after prior feedback exposure (RQ2), and longer-term transfer after both groups had experienced at least one feedback cycle (RQ3).}
  \label{fig:four-stage-workflow}
\end{figure}

We organized the study as a staged intervention design across four phases, as illustrated in Figure~\ref{fig:four-stage-workflow}. This arrangement supports three complementary contrasts: (1) within-round revision after feedback (RQ1), (2) first-submission quality on a new task after prior feedback exposure (RQ2), and (3) longer-term transfer after both groups had experienced feedback at least once (RQ3).

\begin{itemize}
  \item \textbf{Stage 1: Baseline (APP1).} Both Group A and Group B tested APP1 and submitted initial reports. These submissions established the baseline performance of the two groups on the same task.

  \item \textbf{Stage 2: Intervention vs. Control (APP1 vs. APP2).} Group A received agent-generated feedback on its APP1 reports and revised/resubmitted those reports. Group B received no feedback at this point and instead moved to APP2, where it submitted initial reports. This stage both supports the within-round revision analysis for Group A and establishes a no-feedback APP2 baseline for Group B.

  \item \textbf{Stage 3: New Task / Delayed Intervention (APP2).} After prior feedback exposure on APP1, Group A moved to APP2 and submitted initial reports. Group B then received agent-generated feedback on the APP2 reports it had submitted in Stage 2 and revised/resubmitted them. Comparing Group A's initial APP2 reports with Group B's initial APP2 reports from Stage 2 isolates the effect of prior feedback on first-submission quality in a new task.

  \item \textbf{Stage 4: Skill Transfer (APP3).} Finally, both groups tested APP3 and submitted initial reports. Because both groups had now experienced at least one feedback cycle, this stage supports the evaluation of longer-term transfer on a distinct application.
\end{itemize}

\subsection{Post-task Questionnaire Design and Administration}
\label{sec:questionnaire-method}

To complement the artifact-based evidence collected in the four-stage study, we administered a post-task questionnaire after all stages had been completed. The questionnaire served two goals. First, it provided a clearer characterization of the participant pool beyond platform records alone, including prior crowdsourced-testing experience, training background, technical experience, and familiarity with the three application types. Second, it collected complementary evidence about how participants perceived and used the agent-generated feedback, as well as whether external AI tools might have influenced the observed outcomes.

Table~\ref{tab:questionnaire_overview} summarizes the structure of the questionnaire. Rather than functioning as a second primary evaluation instrument, the questionnaire was designed to contextualize the participant pool, identify possible confounds, and provide additional evidence about feedback understanding, trust, adoption, and transfer.

\begin{table}[t]
  \caption{Overview of the post-task questionnaire}
  \label{tab:questionnaire_overview}
  \footnotesize
  \centering
  \setlength{\tabcolsep}{3pt}
  \renewcommand{\arraystretch}{1.08}
  \begin{tabularx}{\linewidth}{@{}>{\raggedright\arraybackslash}p{0.17\linewidth} >{\raggedright\arraybackslash}X >{\raggedright\arraybackslash}p{0.17\linewidth} >{\raggedright\arraybackslash}p{0.23\linewidth}@{}}
    \toprule
    \textbf{Component} & \textbf{Content} & \textbf{Response format} & \textbf{Role in this paper} \\
    \midrule
    Participant background & Prior crowdsourced-testing experience, training background, technical/project experience, and familiarity with the application types & Multiple-choice \& Likert-style & Participant characterization (\S\ref{sec4}) \\[2pt]
    AI-tool usage & General AI-tool habits and possible use of external AI tools during the tasks, including purposes and perceived influence & Multiple-choice, multi-select \& ordinal & Confound checking (\S\ref{sec:questionnaire-findings}) \\[2pt]
    Prior exposure to automated feedback & Prior experience with automated scoring/feedback and habits of revising based on feedback & Multiple-choice \& Likert-style & Context for interpreting feedback uptake \\[2pt]
    Task engagement \& requirement habits & Self-reported engagement, attention to requirement documents, and function-point decomposition habits & Likert-style & Context for interpreting adequacy and transfer \\[2pt]
    Perceived feedback effects & Perceived clarity, trustworthiness, actionability, adoption, and transfer of the agent-generated feedback & Likert-style & Complementary evidence (\S\ref{sec:questionnaire-findings}) \\[2pt]
    Semi-open responses & Most helpful parts, confusing parts, transferred practices, and desired improvements & Semi-open & Thematic analysis (\S\ref{sec:questionnaire-findings}) \\
    \bottomrule
  \end{tabularx}
\end{table}

A total of 17 of the 20 participants completed the questionnaire, yielding a response rate of 85.0\%. The respondent subset included 8 of the 10 testers in Group A and 9 of the 10 testers in Group B, so both study groups remained represented in the questionnaire evidence. We use the questionnaire in two different ways in this paper. In Section~\ref{sec4}, it is used only for participant characterization, so that the respondent subset can be described more precisely. In Section~\ref{sec:questionnaire-findings}, it is used as complementary evidence to help interpret the main artifact-based results, especially with respect to feedback understanding, trust, adoption, transfer, and possible external-AI confounds.

For the structured items, we report descriptive statistics such as distributions, means, medians, and agreement rates when appropriate. For the semi-open items, we use lightweight thematic grouping to summarize recurring patterns in participant responses. These questionnaire findings are not treated as replacements for the primary outcome measures derived from report artifacts; rather, they serve as supplementary evidence for interpreting the mechanisms, boundaries, and practical implications of the observed improvements.

\subsection{Measures and Analysis}
To answer the three RQs, we collected all artifacts submitted by the 20 testers across the four stages, including initial testcase documents, defect reports, revised submissions, and the feedback logs generated during the intervention stages. The primary objective measures in this study were the Textuality and Adequacy outputs produced by the assessment backbone described in Section~\ref{sec3}. These outputs served as the main artifact-based indicators of report quality. Competitiveness was not analyzed as an intervention measure for the reason noted above: it is inherently post-hoc and depends on the complete report set.

The analyses were organized according to the staged comparisons underlying each RQ. For RQ1, we compared reports before and after feedback within the intervention stages (Stage 1 to Stage 2 for Group A on APP1; Stage 2 to Stage 3 for Group B on APP2) to evaluate immediate revision gains. For RQ2, we compared the initial APP2 submissions from Group A, which had prior feedback exposure on APP1, with the initial APP2 submissions from Group B collected before Group B received any APP2 feedback. This isolates the effect of prior feedback on first-submission quality in a new task. For RQ3, we compared the initial baseline reports on APP1 with the later initial reports on APP3 after both groups had experienced at least one feedback cycle, thereby evaluating longer-term transfer to a different application.

Questionnaire data were analyzed descriptively because of the small respondent subset. Structured items were summarized with response distributions and aggregate statistics, while semi-open responses were grouped into lightweight themes. Throughout the paper, questionnaire findings are used as complementary evidence for participant characterization, confounding-factor assessment, and interpretation of feedback perception and transfer, rather than as substitutes for the primary artifact-based outcome measures.

\section{Results}
\label{sec5}
\subsection{RQ1: Effectiveness of Agent-Human Interaction on Report Revision}

\textbf{Motivation.} Prior work on \emph{LLM-as-a-Judge} has primarily established assessor consistency and efficiency. In a real crowdsourced testing workflow, the practical question is whether agent-generated, actionable feedback can immediately improve report quality \emph{within the same round} without changing task goals or time budgets. RQ1 therefore evaluates the \emph{immediate} value of Agent--Human interaction: can feedback alone yield measurable gains in textual quality and adequacy, thereby reducing developer review cost and preparing testers for subsequent tasks?

\begin{table}[h]
  \caption{RQ1 baseline per-tester scores (Stage 1, split by group, on APP1)}
  \label{tab:rq1_baseline_split}
  \small
  \centering
  \begin{threeparttable}
  \begin{tabular}{c c c c c c c c c c}
    \toprule
    \multicolumn{5}{c}{\textbf{Group A}} & \multicolumn{5}{c}{\textbf{Group B}} \\
    \cmidrule(lr){1-5} \cmidrule(lr){6-10}
    \textbf{TID} & \textbf{TC} & \textbf{DE} & \textbf{TXT} & \textbf{ADQ} & \textbf{TID} & \textbf{TC} & \textbf{DE} & \textbf{TXT} & \textbf{ADQ} \\
    \midrule
   1 & 90.62 & 85.26 & 87.94 & 88.57 & 11 & 76.88 & 86.32 & 81.60 & 68.57 \\
   2 & 89.37 & 82.63 & 86.00 & 91.43 & 12 & 86.25 & 86.31 & 86.28 & 85.71 \\
   3 & 88.12 & 85.79 & 86.96 & 34.29 & 13 & 71.88 & 83.16 & 77.52 & 82.86 \\
   4 & 89.38 & 85.79 & 87.58 & \maxv{100.00} & 14 & 91.25 & 88.42 & 89.83 & \maxv{100.00} \\
   5 & 74.38 & 84.74 & 79.56 & 91.43 & 15 & 89.38 & 87.37 & 88.37 & 85.71 \\
   6 & 88.12 & 77.89 & 83.01 & 91.43 & 16 & \maxv{94.38} & 89.47 & 91.92 & 91.43 \\
   7 & 89.37 & 87.37 & \maxv{88.37} & 80.00 & 17 & 70.63 & 73.16 & 71.89 & 91.43 \\
   8 & \maxv{96.25} & \minv{71.05} & 83.65 & 74.29 & 18 & \minv{64.37} & \minv{64.47} & \minv{64.42} & \minv{62.86} \\
   9 & \minv{48.12} & 71.58 & \minv{59.85} & 91.43 & 19 & 92.50 & \maxv{91.58} & \maxv{92.04} & 91.43 \\
   10 & 81.87 & \maxv{88.60} & 85.23 & \minv{11.43} & 20 & 85.62 & 86.84 & 86.23 & 88.57 \\
    \midrule
    Ave\textsubscript{A} & 83.56 & 82.07 & 82.82 & 75.43 & Ave\textsubscript{B} & 82.31 & 83.71 & 83.01 & 84.86 \\
    \bottomrule
  \end{tabular}
  \begin{tablenotes}\footnotesize
    \item Abbreviations: TC = testcase textuality score; DE = defect textuality score; TXT = textuality score; ADQ = adequacy coverage percent.
  \end{tablenotes}
  \end{threeparttable}
\end{table}

\textbf{Baseline (group comparability).} Although Groups A and B were stratified and paired by historical performance, we explicitly validate that their starting levels on the same application (APP1) are comparable before running the intervention/control contrast. Table~\ref{tab:rq1_baseline_split} reports APP1 baselines. Group means on the three textual metrics are close (A: TC=83.56, DE=82.07, TXT=82.82; B: TC=82.31, DE=83.71, TXT=83.01), with absolute differences roughly 1--2 points, providing no evidence of a large systematic gap. For adequacy (ADQ), A's mean is 75.43 and B's is 84.86, slightly lower for A.

Inspecting the range of scores clarifies individual strengths and simultaneously confirms the significant quality variance typical of crowdsourced testing environments. These results show that some testers, likely those with more experience, consistently produce high-quality reports (e.g., TID 16 and 19 in Group B), while others submit work of a much lower quality. These results also highlights specific strengths, such as TID 7 (A) excelling in textual clarity/structure (TXT) and TID 8 (A) in test-case design (TC). Furthermore, Group A's TID 10 presents a pronounced ADQ outlier (11.43); a plausible cause, corroborated by qualitative inspection, is a misunderstanding of the requirement hierarchy and an overly narrow coverage strategy—over-focusing on a single path and UI-flow reuse instead of decomposing and mapping test cases to leaf-level functional points—leading to an underestimated coverage score.To reduce distortion from this outlier, we recompute a robust ADQ mean for Group A after excluding TID 10, which raises A's ADQ mean from 75.43 to 82.54. The difference from Group B (84.86) shrinks to about 2.32 points. Together with the already-close textual means, this supports that Groups A and B are comparable at baseline and appropriate for the subsequent intervention (A) vs. non-intervention (B) contrast.

\begin{table}[H]
  \caption{RQ1 pre- vs. post-feedback scores per tester (A: APP1 \textrightarrow{} APP1$_e$; B: APP2 \textrightarrow{} APP2$_e$)}
  \label{tab:rq1_per_tester}
  \small
  \centering
  \begin{threeparttable}
  \begin{tabular}{l r r r r r r r r r}
    \toprule
    \multirow{2}{*}{\textbf{APP}} & \multirow{2}{*}{\textbf{TID}} & \multicolumn{4}{c}{\textbf{Before enhancement}} & \multicolumn{4}{c}{\textbf{After enhancement}} \\
    \cmidrule(lr){3-6} \cmidrule(lr){7-10}
     &  & \textbf{TC} & \textbf{DE} & \textbf{TXT} & \textbf{ADQ} & \textbf{TC} & \textbf{DE} & \textbf{TXT} & \textbf{ADQ} \\
    \midrule
    \multirow{10}{*}{app1} & 1 & 90.62 & 85.26 & 87.94 & 88.57 & \dec{81.25} & \inc{89.48} & \dec{85.36} & \inc{100.00} \\
    & 2 & 89.37 & 82.63 & 86.00 & 91.43 & \inc{92.50} & \inc{86.31} & \inc{89.41} & \inc{100.00} \\
    & 3 & 88.12 & 85.79 & 86.96 & 34.29 & \inc{93.75} & \inc{87.89} & \inc{90.82} & \inc{94.29} \\
    & 4 & 89.38 & 85.79 & 87.58 & 100.00 & \inc{95.63} & \inc{91.58} & \inc{93.60} & \same{100.00} \\
    & 5 & 74.38 & 84.74 & 79.56 & 91.43 & \inc{96.25} & \inc{92.11} & \inc{94.18} & \inc{100.00} \\
    & 6 & 88.12 & 77.89 & 83.01 & 91.43 & \inc{95.63} & \inc{89.48} & \inc{92.55} & \inc{100.00} \\
    & 7 & 89.37 & 87.37 & 88.37 & 80.00 & \inc{91.25} & \inc{91.58} & \inc{91.42} & \inc{100.00} \\
    & 8 & 96.25 & 71.05 & 83.65 & 74.29 & \inc{96.88} & \inc{90.00} & \inc{93.44} & \inc{100.00} \\
    & 9 & 48.12 & 71.58 & 59.85 & 91.43 & \inc{96.88} & \inc{87.89} & \inc{92.39} & \inc{94.29} \\
    & 10 & 81.87 & 88.60 & 85.23 & 11.43 & \inc{93.75} & \inc{89.47} & \inc{91.61} & \inc{74.29} \\
    \midrule
    & Ave\textsubscript{A} & 83.56 & 82.07 & 82.82 & 75.43 & \inc{93.38} & \inc{89.58} & \inc{91.48} & \inc{96.29} \\
    \midrule
    \multirow{10}{*}{app2} & 11 & 76.88 & 87.37 & 82.12 & 93.75 & \inc{96.88} & \inc{87.90} & \inc{92.39} & \inc{100.00} \\
    & 12 & 83.75 & 84.74 & 84.24 & 96.88 & \inc{95.00} & \inc{91.06} & \inc{93.03} & \same{96.88} \\
    & 13 & 68.12 & 82.63 & 75.38 & 87.50 & \inc{95.00} & \inc{91.58} & \inc{93.29} & \inc{100.00} \\
    & 14 & 85.62 & 88.94 & 87.28 & 93.75 & \inc{96.88} & \inc{92.11} & \inc{94.50} & \inc{100.00} \\
    & 15 & 91.25 & 88.95 & 90.10 & 96.88 & \inc{93.12} & \inc{91.05} & \inc{92.09} & \inc{100.00} \\
    & 16 & 94.38 & 89.47 & 91.92 & 96.88 & \inc{96.25} & \same{89.47} & \inc{92.86} & \inc{100.00} \\
    & 17 & 62.50 & 68.42 & 65.46 & 84.38 & \inc{95.00} & \inc{91.58} & \inc{93.29} & \inc{96.88} \\
    & 18 & 44.38 & 37.89 & 41.13 & 87.50 & \inc{72.50} & \inc{41.05} & \inc{56.78} & \inc{100.00} \\
    & 19 & 86.87 & 91.05 & 88.96 & 96.88 & \inc{94.38} & \inc{91.58} & \inc{92.98} & \inc{100.00} \\
    & 20 & 86.87 & 80.53 & 83.70 & 100.00 & \inc{96.25} & \inc{89.48} & \inc{92.87} & \same{100.00} \\
    \midrule
    & Ave\textsubscript{B} & 78.06 & 80.00 & 79.03 & 93.44 & \inc{93.13} & \inc{85.69} & \inc{89.41} & \inc{99.38} \\
    \bottomrule
  \end{tabular}
  \begin{tablenotes}\footnotesize
    \item Cell shading: \textcolor{green!50!black}{green} = post improvement; \textcolor{red!70!black}{red} = post decrease; equal = gray.
  \end{tablenotes}
  \end{threeparttable}
\end{table}

\textbf{Results.} Reading the pre/post feedback scores in Table~\ref{tab:rq1_per_tester}, we observe consistent, within-round gains after agent feedback. For Group A on APP1 (app1 \textrightarrow{} app1$_e$), 9/10 testers increased TC and TXT (TID 1 shows a small TC/TXT trade-off), and 8/10 increased ADQ while one remained at ceiling (TID 4: 100\%). Notably, 7/10 testers reached ADQ=100\% after revision (TIDs 1,2,4,5,6,7,8), and two additional testers approached near-ceiling (TIDs 3 and 9 at 94.29\%). For Group B on APP2 (app2 \textrightarrow{} app2$_e$), gains are similarly broad: 10/10 increased TC and TXT, 9/10 increased DE (TID 16 unchanged), and 9/10 increased ADQ (TID 12 unchanged). Moreover, 8/10 testers reached ADQ=100\% after revision (TIDs 11,13,14,15,16,18,19,20), with the remainder maintaining high coverage (TIDs 12 and 17 at 96.88\%). These patterns indicate that feedback is translated into concrete edits that improve textual clarity/structure and, crucially, close coverage gaps.

\begin{table}[h]
  \caption{RQ1 improvement rates per tester and group average}
  \label{tab:rq1_improve_rates}
  \small
  \centering
  \begin{threeparttable}
  \begin{tabular}{l r r r r r}
    \toprule
    \textbf{GR.} & \textbf{TID} & \textbf{TC\,(\%)} & \textbf{DE\,(\%)} & \textbf{TXT\,(\%)} & \textbf{ADQ\,(\%)} \\
    \midrule
    \multirow{11}{*}{A} & 1 & \dec{-10.34\%} & \inc{+4.95\%} & \dec{-2.93\%} & \inc{+12.91\%} \\
    & 2 & \inc{+3.50\%} & \inc{+4.45\%} & \inc{+3.97\%} & \inc{+9.37\%} \\
    & 3 & \inc{+6.39\%} & \inc{+2.45\%} & \inc{+4.44\%} & \inc{+174.98\%} \\
    & 4 & \inc{+6.99\%} & \inc{+6.75\%} & \inc{+6.87\%} & \same{0.00\%} \\
    & 5 & \inc{+29.40\%} & \inc{+8.70\%} & \inc{+18.38\%} & \inc{+9.37\%} \\
    & 6 & \inc{+8.52\%} & \inc{+14.88\%} & \inc{+11.49\%} & \inc{+9.37\%} \\
    & 7 & \inc{+2.10\%} & \inc{+4.82\%} & \inc{+3.45\%} & \inc{+25.00\%} \\
    & 8 & \inc{+0.65\%} & \inc{+26.67\%} & \inc{+11.70\%} & \inc{+34.61\%} \\
    & 9 & \inc{+101.33\%} & \inc{+22.79\%} & \inc{+54.37\%} & \inc{+3.13\%} \\
    & 10 & \inc{+14.51\%} & \inc{+0.98\%} & \inc{+7.49\%} & \inc{+549.96\%} \\
    \midrule
    & Ave\textsubscript{A} & \inc{+11.75\%} & \inc{+9.15\%} & \inc{+10.46\%} & \inc{+27.65\%} \\
    \midrule
    \multirow{11}{*}{B} & 11 & \inc{+26.01\%} & \inc{+0.61\%} & \inc{+12.51\%} & \inc{+6.67\%} \\
    & 12 & \inc{+13.43\%} & \inc{+7.46\%} & \inc{+10.43\%} & \same{0.00\%} \\
    & 13 & \inc{+39.46\%} & \inc{+10.83\%} & \inc{+23.76\%} & \inc{+14.29\%} \\
    & 14 & \inc{+13.15\%} & \inc{+3.56\%} & \inc{+8.27\%} & \inc{+6.67\%} \\
    & 15 & \inc{+2.05\%} & \inc{+2.36\%} & \inc{+2.21\%} & \inc{+3.22\%} \\
    & 16 & \inc{+1.98\%} & \same{0.00\%} & \inc{+1.02\%} & \inc{+3.22\%} \\
    & 17 & \inc{+52.00\%} & \inc{+33.85\%} & \inc{+42.51\%} & \inc{+14.81\%} \\
    & 18 & \inc{+63.36\%} & \inc{+8.34\%} & \inc{+38.05\%} & \inc{+14.29\%} \\
    & 19 & \inc{+8.65\%} & \inc{+0.58\%} & \inc{+4.52\%} & \inc{+3.22\%} \\
    & 20 & \inc{+10.80\%} & \inc{+11.11\%} & \inc{+10.96\%} & \same{0.00\%} \\
    \midrule
    & Ave\textsubscript{B} & \inc{+19.30\%} & \inc{+7.11\%} & \inc{+13.13\%} & \inc{+6.35\%} \\
    \bottomrule
  \end{tabular}
  \begin{tablenotes}\footnotesize
    \item Improvement rate formula: (post $-$ pre) $\div$ pre. Positive = improvement (green), negative = decrease (red), zero = equal (gray). For \emph{Ave} rows, group averages are computed as (mean(post) $-$ mean(pre)) $\div$ mean(pre) across testers.
  \end{tablenotes}
  \end{threeparttable}
\end{table}

As shown in Table~\ref{tab:rq1_improve_rates}, group averages corroborate these observations at the improvement-rate level. On average, Group A achieves +11.75\% (TC), +9.15\% (DE), +10.46\% (TXT), and a large +27.65\% (ADQ), reflecting strong coverage consolidation enabled by adequacy-oriented feedback (requirement-tree mapping and uncovered-point prompts). Group B shows +19.30\% (TC), +7.11\% (DE), +13.13\% (TXT), and +6.35\% (ADQ). Per-tester rates further show that (i) nearly all testers have strictly positive gains across textual metrics, with a few ties/plateaus (e.g., DE for TID 16, ADQ for TID 12); (ii) testers with lower baselines (e.g., TID 9 in A, TID 18 in B) tend to exhibit larger relative jumps, consistent with a headroom effect; and (iii) adequacy improves most where the baseline headroom is largest (A's ADQ mean was lower at baseline), while B's adequacy gains are modest but still positive/near-ceiling for most testers. Altogether, the per-metric distributions and group means present a coherent picture: agent feedback yields immediate, measurable improvements in both textual quality and functional coverage, with especially pronounced effects on coverage when headroom exists.

\begin{tcolorbox}[colback=gray!10,colframe=gray!50,arc=2mm,boxrule=0.5pt]
\textbf{Findings.} (i) Groups A and B are comparable at baseline (after excluding one ADQ outlier in A), satisfying the prerequisite for an intervention/control contrast. (ii) Agent feedback produces immediate, measurable gains within the same round: most testers improve TC/DE/TXT and many close adequacy gaps, with numerous cases reaching ADQ=100\%. (iii) Gains are larger where baseline headroom is greater (e.g., A's ADQ), whereas near-ceiling metrics show smaller but still positive deltas. (iv) The improvements are robust across apps and testers; occasional small trade-offs do not change the overall positive direction.
\end{tcolorbox}

\subsection{RQ2: Impact of Agent Integration on New Task Performance}

\textbf{Motivation.} RQ2 evaluates whether agent feedback \emph{transfers} to a new application (APP2) without any additional guidance, i.e., whether testers who previously received feedback (Group A) produce higher-quality \emph{first submissions} on a new task than testers without prior feedback (Group B). We therefore compare Group A (post-feedback, app2$_{\mathrm{f}}$) against Group B (no prior feedback, app2) on their initial APP2 reports.

\begin{table}[h]
  \caption{Initial scores on APP2: Group A (post-feedback, app2$_{\mathrm{f}}$) vs Group B (no prior feedback, app2)}
  \label{tab:rq2_initial_split}
  \small
  \centering
  \begin{tabular}{r r r r r r r r r r}
    \toprule
    \multicolumn{5}{c}{\textbf{Group A (app2$_{\mathrm{f}}$)}} & \multicolumn{5}{c}{\textbf{Group B (app2)}} \\
    \cmidrule(lr){1-5} \cmidrule(lr){6-10}
    \textbf{TID} & \textbf{TC} & \textbf{DE} & \textbf{TXT} & \textbf{ADQ} & \textbf{TID} & \textbf{TC} & \textbf{DE} & \textbf{TXT} & \textbf{ADQ} \\
    \midrule
    1 & 86.25 & 87.90 & 87.07 & 87.50 & 11 & 76.88 & 87.37 & 82.12 & 93.75 \\
    2 & 88.75 & 81.05 & 84.90 & 96.88 & 12 & 83.75 & 84.74 & 84.24 & 96.88 \\
    3 & 88.12 & 83.68 & 85.90 & 93.75 & 13 & 68.12 & 82.63 & 75.38 & 87.50 \\
    4 & 95.00 & 91.05 & 93.03 & 90.62 & 14 & 85.62 & 88.94 & 87.28 & 93.75 \\
    5 & 93.75 & 91.58 & 92.67 & 90.62 & 15 & 91.25 & 88.95 & 90.10 & 96.88 \\
    6 & 94.38 & 92.11 & 93.24 & 96.88 & 16 & 94.38 & 89.47 & 91.92 & 96.88 \\
    7 & 91.88 & 80.53 & 86.20 & 100.00 & 17 & 62.50 & 68.42 & 65.46 & 84.38 \\
    8 & 77.50 & 86.32 & 81.91 & 100.00 & 18 & 44.38 & 37.89 & 41.13 & 87.50 \\
    9 & 92.50 & 89.47 & 90.98 & 90.62 & 19 & 86.87 & 91.05 & 88.96 & 96.88 \\
    10 & 86.87 & 72.10 & 79.49 & 53.12 & 20 & 86.87 & 80.53 & 83.70 & 100.00 \\
    \midrule
    Ave\textsubscript{A} &\textbf{89.50} & \textbf{85.58} & \textbf{87.54} & 90.00 & Ave\textsubscript{B} & 78.06 & 80.00 & 79.03 & \textbf{93.44} \\
    \bottomrule
  \end{tabular}
\end{table}
\textbf{Results.} Two complementary views lead to the same conclusion. First, the per-tester split in Table~\ref{tab:rq2_initial_split} shows that Group A's first submissions on APP2 are consistently stronger in textual quality. Second, the group-level summary in Table~\ref{tab:rq2_group_compare} quantifies this advantage: Group A outperforms Group B on TC (+11.44), DE (+5.58), and TXT (+8.51), indicating that prior feedback generalizes to structure, clarity, and defect write-ups on a new app. For adequacy (ADQ), the headline mean is slightly lower for Group A (90.00) than for Group B (93.44). A closer look reveals one very low-coverage case in Group A (TID~10, ADQ=53.12) that depresses the mean; removing this single outlier yields a robust ADQ mean of $94.10\ (= (900.00-53.12)/9)$, which surpasses Group B's 93.44. Notably, multiple A testers already reach ADQ=100.00 (e.g., TIDs 7 and 8), while most others are near ceiling. 

Zooming in on the recurring outlier (A: TID~10) across stages helps triangulate the root cause. In the APP1 baseline (Table~\ref{tab:rq1_baseline_split}), this tester's ADQ was extremely low (11.43), which we attributed to misunderstanding the requirement hierarchy and adopting a narrow coverage strategy. After receiving adequacy-focused feedback on APP1, the tester's ADQ rose to 74.29 in the revision (Table~\ref{tab:rq1_per_tester}, app1$_e$), showing that the feedback was actionable. On the new APP2 task, however, the tester again under-estimated coverage (ADQ=53.12; Table~\ref{tab:rq2_initial_split}) despite reasonable textual scores (e.g., TC=86.87, TXT=79.49). Consistent with this diagnosis, their APP2 cases conflate composite parent requirements with single functional points, omit sibling branches under common parents, and reuse step templates across distinct flows—choices that boost textual coherence but bypass leaf-level enumeration and mapping, thereby systematically undercounting coverage. For some testers, this pattern suggests that while feedback is actionable, a single intervention is insufficient to override ingrained testing habits. This highlights the importance of integrating assessment agents into the crowdsourcing workflow, where multiple rounds of feedback may be necessary to reinforce new practices and achieve lasting skill improvement.

\begin{table}[h]
  \caption{RQ2 group-level comparison on APP2 (means and differences)}
  \label{tab:rq2_group_compare}
  \small
  \centering
  \begin{tabular}{l l c c c}
    \toprule
    \textbf{Code} & \textbf{Metric} & \textbf{Group A (app2$_{\mathrm{f}}$)} & \textbf{Group B (app2)} & \textbf{$\Delta$ (A$-$B)}\\
    \midrule
    \textbf{TC} & Testcase Textuality Score & \inc{89.50} & \dec{78.06} & \inc{+11.44} \\
    \textbf{DE} & Defect Textuality Score & \inc{85.58} & \dec{80.00} & \inc{+5.58} \\
    \textbf{TXT} & Textuality Score & \inc{87.54} & \dec{79.03} & \inc{+8.51} \\
    \textbf{ADQ} & Adequacy Coverage Percent & \dec{90.00} & \inc{93.44} & \dec{-3.44} \\
    \textbf{ADQ*} & Adequacy (w/o outlier) & \inc{94.10} & \dec{93.44} & \inc{+0.66} \\
    \bottomrule
  \end{tabular}
\end{table}

The results in Table~\ref{tab:rq2_group_compare} highlight a key pattern in skill transfer. Group A, having received prior feedback, demonstrates substantially higher scores across all textual metrics (TC: +11.44, DE: +5.58, TXT: +8.51) compared to Group B. This indicates that the principles of writing clear, structured, and complete reports are generalizable and were successfully transferred to the new task. In contrast, the headline Adequacy (ADQ) score for Group A is slightly lower, showing a negative delta of -3.44. However, as noted previously, this is attributable to a single outlier. The `ADQ*` row shows that after removing this case, Group A's robust mean (94.10) slightly surpasses Group B's (93.44), reversing the delta to +0.66.

The minimal difference in ADQ scores, with both groups achieving high coverage, is likely due to the nature of the test application itself. APP2 is an online shopping platform, a type of application with which most testers have significant prior experience from everyday use. This familiarity likely allowed them to more intuitively and accurately deconstruct the functional requirements, regardless of whether they had received prior agent feedback. In contrast to a more specialized or complex management system, where a systematic approach is critical, the common user workflows of an e-commerce site are well-understood. This 'familiarity effect' appears to have leveled the playing field for the adequacy dimension on this specific task, leading to high baseline performance from both groups and masking some of the transferable impact of the agent's feedback on coverage strategy.In contrast, the textuality dimension exhibits a clear and significant improvement for the feedback-exposed group.Qualitative inspection indicates that these gains stem from behaviors directly reinforced by the agent feedback: more explicit preconditions and oracles, tighter step--expected result alignment, more consistent terminology aligned with requirement phrasing, and pruning of extraneous narrative, rather than from app familiarity alone. In short, while adequacy tends to saturate on a familiar e-commerce app, writing quality continues to benefit from the learned scaffolding, yielding significantly higher textuality socre on the new task.

\begin{tcolorbox}[colback=gray!10,colframe=gray!50,arc=2mm,boxrule=0.5pt]
\textbf{Findings.} (i) Agent feedback on textual quality transfers effectively to new tasks. Group A significantly outperformed the control group on TC, DE, and TXT, demonstrating that universal writing skills are readily generalized. (ii) Adequacy transfer appears muted on APP2 due to a \emph{familiarity effect}: common e-commerce flows let both groups reach near-ceiling coverage, compressing between-group differences. After removing one low-coverage outlier, Group A's robust ADQ mean slightly exceeds Group B, indicating transferable coverage skills that are partly masked by domain familiarity. 
\end{tcolorbox}

\subsection{RQ3: Skill Transferability to a New Application}

\textbf{Motivation.} RQ3 examines whether capabilities learned through agent feedback \emph{extend over time and across tasks}. Specifically, after both groups have experienced at least one feedback cycle in the prior stages, we assess whether testers demonstrate higher-quality first submissions on a distinct application (APP3) compared to their initial baseline on APP1, indicating durable, transferable skill gains.

\begin{table}[h]
  \caption{Per-tester scores on APP1 (baseline) vs APP3 (transfer)}
  \label{tab:rq3_split}
  \small
  \centering
  \begin{tabular}{c c c c c c c c c c}
    \toprule
    \multicolumn{5}{c}{\textbf{APP1 (baseline)}} & \multicolumn{5}{c}{\textbf{APP3 (transfer)}} \\
    \cmidrule(lr){1-5} \cmidrule(lr){6-10}
    \textbf{TID} & \textbf{TC} & \textbf{DE} & \textbf{TXT} & \textbf{ADQ} & \textbf{TID} & \textbf{TC} & \textbf{DE} & \textbf{TXT} & \textbf{ADQ} \\
    \midrule
    1 & \inc{90.62} & \inc{85.26} & \inc{87.94} & \inc{88.57} & 1 & \dec{85.63} & \dec{76.32} & \dec{80.97} & \dec{68.42} \\
    2 & \inc{89.37} & \inc{82.63} & \inc{86.00} & \inc{91.43} & 2 & \dec{88.75} & \dec{71.58} & \dec{80.16} & \dec{91.23} \\
    3 & \inc{88.12} & \inc{85.79} & \inc{86.96} & \dec{34.29} & 3 & \dec{85.63} & \dec{78.42} & \dec{82.02} & \inc{89.47} \\
    4 & \inc{89.38} & \dec{85.79} & \dec{87.58} & \inc{100.00} & 4 & \dec{89.37} & \inc{91.05} & \inc{90.21} & \dec{84.21} \\
    5 & \dec{74.38} & \dec{84.74} & \dec{79.56} & \dec{91.43} & 5 & \inc{93.13} & \inc{91.58} & \inc{92.35} & \inc{96.49} \\
    6 & \inc{88.12} & \dec{77.89} & \dec{83.01} & \inc{91.43} & 6 & \dec{84.37} & \inc{83.68} & \inc{84.03} & \dec{91.23} \\
    7 & \dec{89.37} & \inc{87.37} & \dec{88.37} & \dec{80.00} & 7 & \inc{96.88} & \dec{83.16} & \inc{90.02} & \inc{91.23} \\
    8 & \inc{96.25} & \dec{71.05} & \dec{83.65} & \dec{74.29} & 8 & \dec{90.62} & \inc{88.95} & \inc{89.78} & \inc{89.47} \\
    9 & \dec{48.12} & \dec{71.58} & \dec{59.85} & \inc{91.43} & 9 & \inc{83.75} & \inc{87.89} & \inc{85.82} & \dec{78.95} \\
    10 & \dec{81.87} & \inc{88.60} & \dec{85.23} & \dec{11.43} & 10 & \inc{86.87} & \dec{86.84} & \inc{86.86} & \inc{47.37} \\
    11 & \dec{76.88} & \dec{86.32} & \dec{81.60} & \dec{68.57} & 11 & \inc{94.38} & \inc{91.58} & \inc{92.98} & \inc{96.49} \\
    12 & \dec{86.25} & \inc{86.31} & \dec{86.28} & \dec{85.71} & 12 & \inc{93.13} & \dec{85.79} & \inc{89.46} & \inc{94.74} \\
    13 & \dec{71.88} & \dec{83.16} & \dec{77.52} & \dec{82.86} & 13 & \inc{88.75} & \inc{89.47} & \inc{89.11} & \inc{98.25} \\
    14 & \dec{91.25} & \dec{88.42} & \dec{89.83} & \inc{100.00} & 14 & \inc{91.87} & \inc{91.06} & \inc{91.47} & \dec{92.98} \\
    15 & \dec{89.38} & \inc{87.37} & \dec{88.37} & \dec{85.71} & 15 & \inc{90.62} & \dec{86.32} & \inc{88.47} & \inc{94.74} \\
    16 & \same{94.38} & \dec{89.47} & \dec{91.92} & \dec{91.43} & 16 & \same{94.38} & \inc{92.11} & \inc{93.24} & \inc{100.00} \\
    17 & \dec{70.63} & \dec{73.16} & \dec{71.89} & \dec{91.43} & 17 & \inc{91.25} & \inc{81.05} & \inc{86.15} & \inc{92.98} \\
    18 & \dec{64.37} & \dec{64.47} & \dec{64.42} & \dec{62.86} & 18 & \inc{86.25} & \inc{73.16} & \inc{79.71} & \inc{66.67} \\
    19 & \dec{92.50} & \inc{91.58} & \inc{92.04} & \dec{91.43} & 19 & \inc{93.75} & \dec{90.00} & \dec{91.88} & \inc{92.98} \\
    20 & \dec{85.62} & \dec{86.84} & \dec{86.23} & \dec{88.57} & 20 & \inc{95.63} & \inc{90.53} & \inc{93.08} & \inc{92.98} \\
    \midrule
    Ave & 82.94 & 82.89 & 82.91 & 80.14 & Ave & 90.25 & 85.53 & 87.89 & 87.54 \\
    \bottomrule
  \end{tabular}
\end{table}

\textbf{Results.} At the individual level (Table~\ref{tab:rq3_split}), most testers improve from APP1 to APP3 across textuality (TC/DE/TXT), and many close coverage gaps. Counting row-wise changes, about two-thirds of testers improve on TC (\~13/20), over half improve on DE (\~12/20), and a large majority improve on TXT (\~17/20). Adequacy (ADQ) rises for most testers (\~15/20), with a handful of decreases typically from near-ceiling baselines. The strongest movers include TID~5 (TC: 74.38\,$\to$\,93.13; TXT: 79.56\,$\to$\,92.35; ADQ: 91.43\,$\to$\,96.49), TID~9 (TC: 48.12\,$\to$\,83.75; DE: 71.58\,$\to$\,87.89; TXT: 59.85\,$\to$\,85.82), TID~11 (all four metrics up, ADQ: 68.57\,$\to$\,96.49), TID~13 (all four metrics up, ADQ: 82.86\,$\to$\,98.25), and TID~16 (TXT: 91.92\,$\to$\,93.24; ADQ: 91.43\,$\to$\,100.00). A small subset shows mixed movements and trade-offs: TID~4 improves DE/TXT but ADQ drops from 100.00\,$\to$\,84.21 (ceiling regression to a high but sub-max level), TID~14 improves textuality with a modest ADQ dip (100.00\,$\to$\,92.98), and TID~7 raises TC/TXT/ADQ while DE softens slightly. A few cases (e.g., TID~1/2) present broader regressions on textual metrics when moving to APP3, consistent with individual adaptation variability; however, these remain exceptions in the aggregate pattern.

\begin{table}[h]
  \caption{Group-level comparison: APP3 (transfer) vs APP1 (baseline)}
  \label{tab:rq3_group_compare}
  \small
  \centering
  \begin{tabular}{l l c c c}
    \toprule
    Code & Metric & APP1 (baseline) & APP3 (transfer) & $\Delta$ (APP3$-$APP1) \\
    \midrule
    TC & testcase textuality score & 82.94 & 90.25 & \inc{+7.31} \\
    DE & defect textuality score & 82.89 & 85.53 & \inc{+2.64} \\
    TXT & textuality score & 82.91 & 87.89 & \inc{+4.98} \\
    ADQ & adequacy coverage percent & 80.14 & 87.54 & \inc{+7.40} \\
    \bottomrule
  \end{tabular}
\end{table}

At the group level (Table~\ref{tab:rq3_group_compare}), averages rise substantially: +7.31 (TC), +2.64 (DE), +4.98 (TXT), and +7.40 (ADQ). The concurrent gains in textuality and adequacy indicate that skills trained via agent feedback persist beyond the intervention window and transfer to a new domain. Put differently, testers not only write clearer, better structured test cases and defect reports on APP3, but also map them more completely to requirement leaf points, with many individuals either attaining or approaching ceiling ADQ. Overall, the group-level pattern remains clearly positive despite a small number of individual regressions.

Still, the mixed per-tester movements in Table~\ref{tab:rq3_split} warrant a more nuanced interpretation. Unlike RQ1, which examines revision within the same task, RQ3 compares first submissions across different applications. Monotonic improvement is therefore not expected for every tester or every metric. A plausible explanation is that transfer is both partial and application-sensitive. Some practices reinforced by the feedback, such as clearer steps, more explicit expected results, and better overall report structure, appear to transfer relatively smoothly. In contrast, deeper habits related to requirement decomposition, coverage planning, and defect-specific reporting may depend more strongly on the functional structure and workflow characteristics of the new application, and thus may transfer less uniformly. Some of the observed dips, especially when APP1 scores were already high, are also consistent with ceiling regression rather than with a substantive loss of capability. Taken together, the mixed movements in Table~\ref{tab:rq3_split} are better understood as reflecting cross-application adaptation costs, ceiling regression, and uneven internalization of different sub-skills, rather than as evidence against longer-term transfer.

\begin{tcolorbox}[colback=gray!10,colframe=gray!50,arc=2mm,boxrule=0.5pt]
\textbf{Findings.} (i) Skills reinforced by agent feedback are durable and transferable: first submissions on APP3 outperform APP1 on both textuality and adequacy. (ii) Gains are broad-based across testers, not driven by a few outliers. (iii) Convergent growth in TC/DE/TXT and ADQ suggests that testers internalize both micro-level writing practices and macro-level coverage planning.
\end{tcolorbox}

\subsection{Complementary Questionnaire Findings}
\label{sec:questionnaire-findings}

To complement the artifact-based results reported in Sections~5.1--5.3, we analyzed the post-task questionnaire completed by 17 of the 20 participants. Overall, the questionnaire results reinforce the main empirical pattern observed in the report artifacts: the agent-generated feedback was generally understandable, acted upon in revision, and carried forward to later tasks. The questionnaire is therefore useful not only for identifying possible external-AI confounds, but also for explaining how the observed quality improvements were achieved in practice.

\begin{figure}[t]
  \centering
  \includegraphics[width=\linewidth]{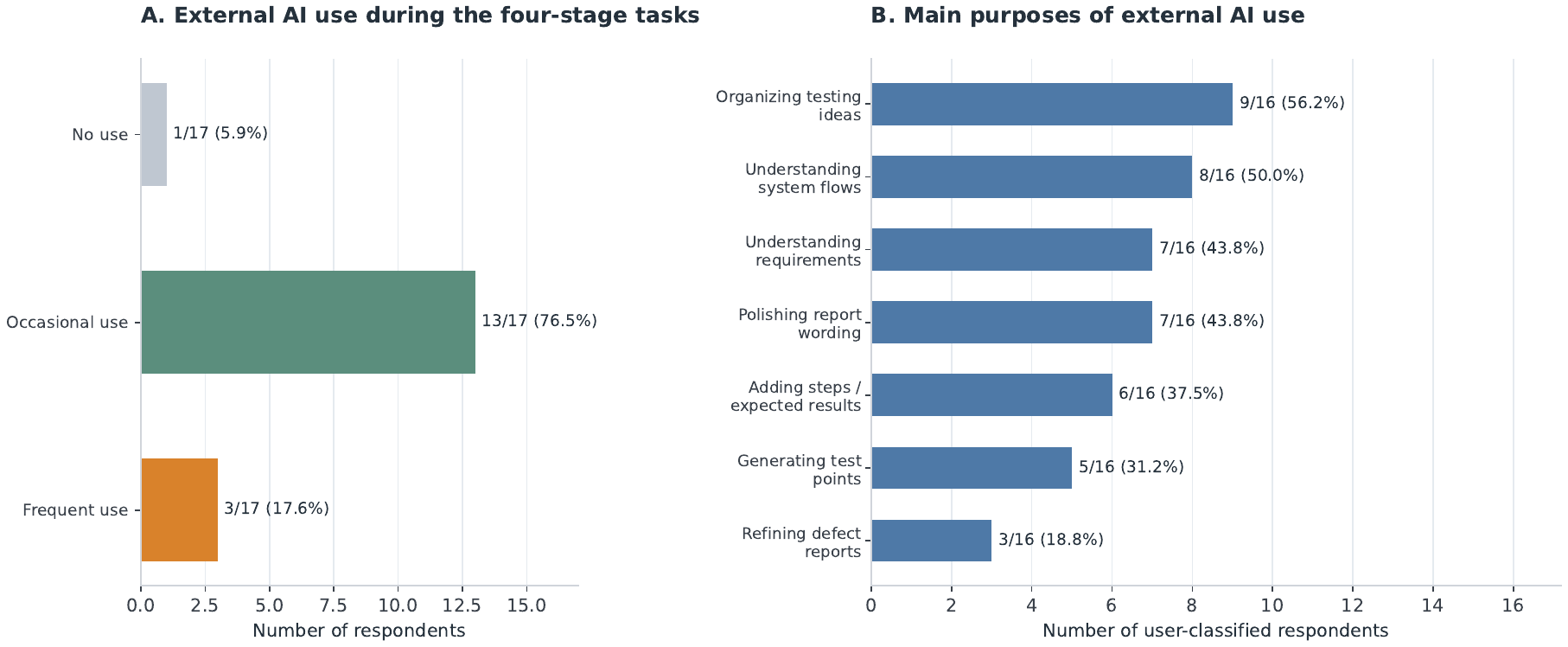}
  \caption{Questionnaire evidence on external AI use during the four-stage tasks. Panel A summarizes the extent of external-AI use, while Panel B shows the most common purposes among the respondents who reported using other AI tools.}
  \label{fig:questionnaire-ai-usage}
\end{figure}

We first examined possible external-AI confounds. As shown in Figure~\ref{fig:questionnaire-ai-usage}, the use of other AI tools was not absent: 16 of the 17 respondents reported at least some use of external AI assistance during the four-stage tasks. However, the pattern is more consistent with partial support than with wholesale outsourcing. Most respondents reported only occasional use for a few parts of the tasks (13/17, 76.5\%), whereas only a small minority reported frequent use across the four stages (3/17, 17.6\%). Among the user-classified respondents, the most common uses were organizing testing ideas, understanding system functionality or page flows, understanding requirements, and polishing testcase/report wording. Only one respondent reported a large influence on the final submission. Thus, external AI use should be treated as a real confound, but the self-reports suggest that it more often played a supportive role in comprehension, planning, and wording than a dominant role in replacing report production.

\begin{figure}[t]
  \centering
  \includegraphics[width=\linewidth]{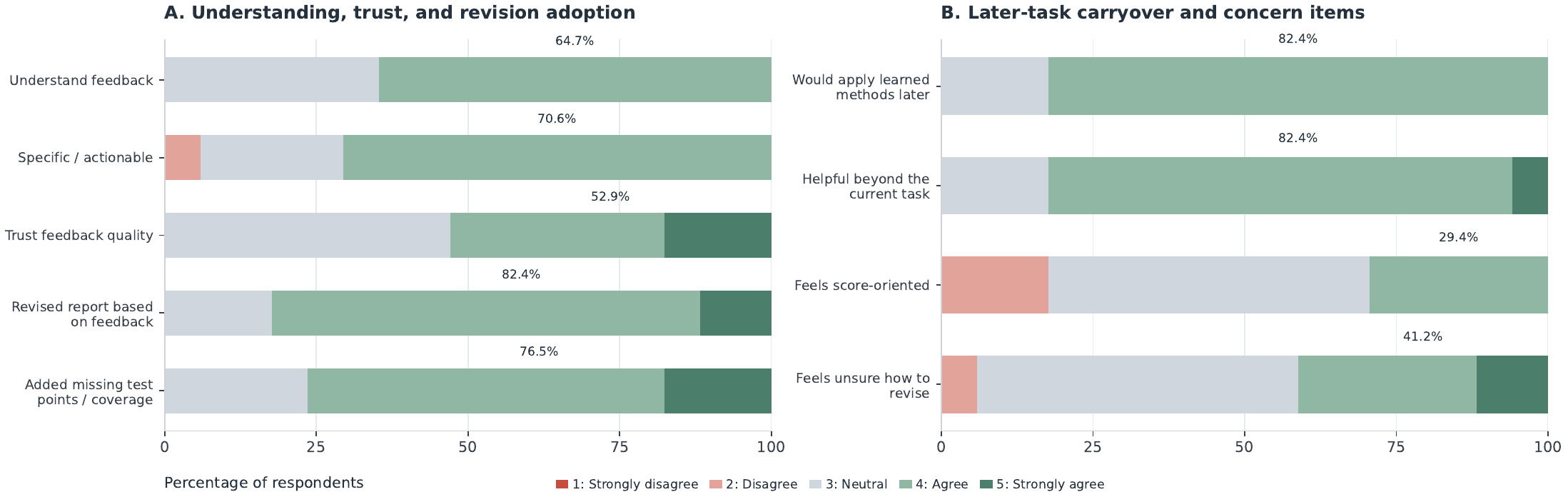}
  \caption{Selected questionnaire items on feedback perception and use. The items are split into two panels for readability. The stacked bars show the full 1--5 response distributions, and the labels above the bars indicate the share of respondents selecting 4--5 on each item.}
  \label{fig:questionnaire-feedback}
\end{figure}

We then examined how the respondent subset perceived and used the agent-generated feedback. Figure~\ref{fig:questionnaire-feedback} shows a broadly positive pattern. Most respondents reported that the feedback was understandable and actionable: 64.7\% agreed that they could understand the feedback, 70.6\% agreed that the suggestions were specific and actionable, and 76.5\% agreed that the feedback helped them notice previously overlooked problems. More importantly, the feedback was not merely read; it was translated into concrete report changes. A large majority agreed that they revised the wording, structure, or details of their reports based on the feedback (82.4\%) and that they added missing test points or requirement coverage (76.5\%). Reported carryover to later tasks was similarly strong: 82.4\% agreed that they would proactively apply the learned methods in later tasks, and the same proportion agreed that the feedback remained useful beyond the immediate task. Taken together, these responses provide complementary evidence that the feedback mechanism was behaviorally consequential for the respondent subset.

\begin{figure}[t]
  \centering
  \includegraphics[width=\linewidth]{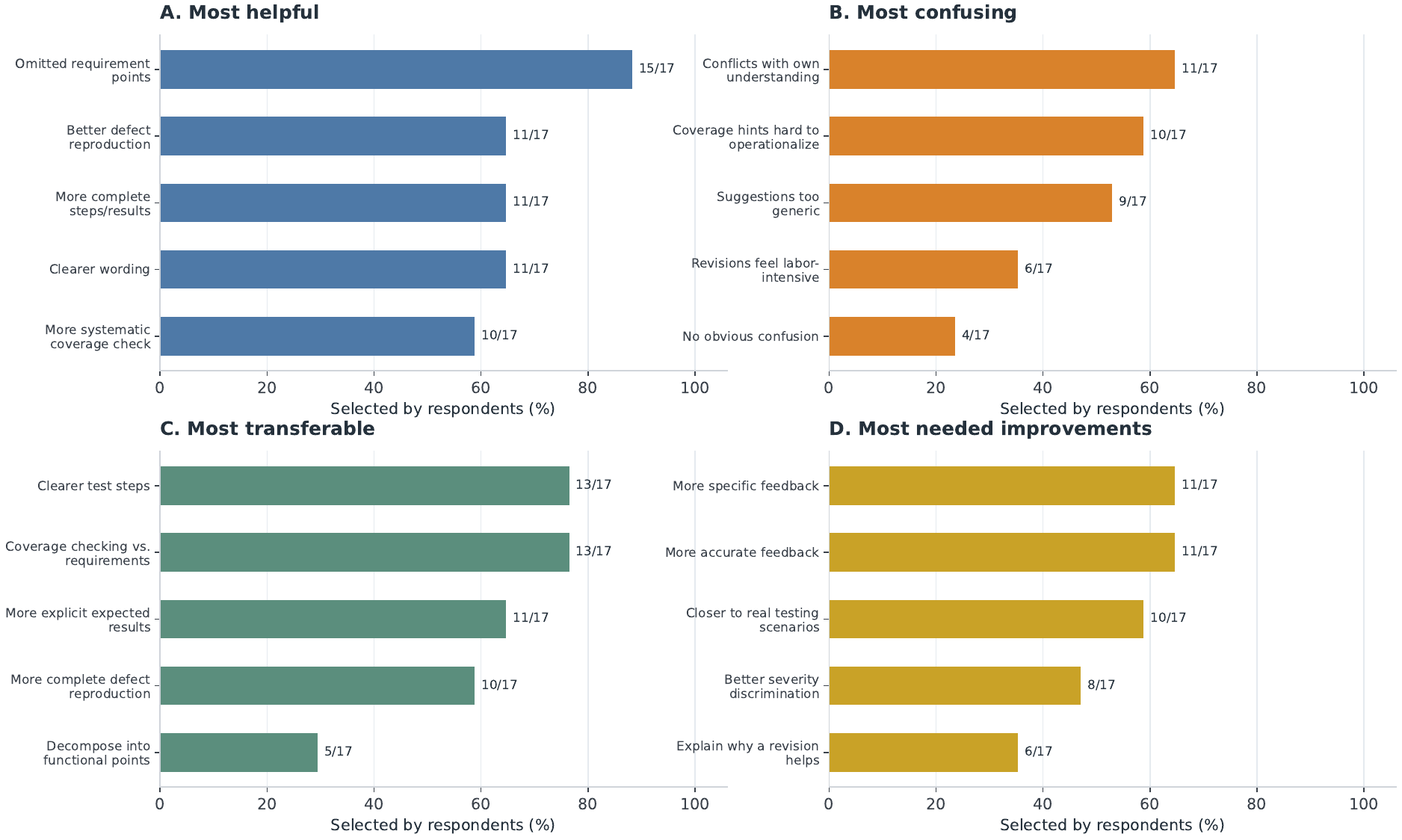}
  \caption{Semi-open questionnaire themes. The four panels summarize what respondents considered most helpful, most confusing, most transferable to later tasks, and most in need of improvement.}
  \label{fig:questionnaire-semi-open}
\end{figure}

The semi-open responses in Figure~\ref{fig:questionnaire-semi-open} help explain what participants appear to have internalized and why the artifact-based results show gains beyond immediate revision. In the ``most helpful'' panel, the dominant selections center on omitted requirement or functional points, better defect reproduction descriptions, and more complete steps or expected results, indicating that participants valued concrete guidance on both coverage and report completeness rather than abstract scoring alone. The ``most transferable'' panel aligns well with the patterns observed in RQ2 and RQ3: respondents most often reported carrying forward clearer test steps, more deliberate requirement-coverage checking, clearer expected results, and more complete defect reproduction information. These themes suggest that the feedback reinforced both micro-level writing practices and broader coverage-planning habits.

At the same time, the questionnaire does not suggest a frictionless mechanism. The ``most confusing'' and ``most needed improvements'' panels show that some respondents felt that parts of the feedback conflicted with their own understanding, that some coverage reminders were hard to operationalize into concrete test points, and that some suggestions were too generic. This pattern is consistent with the structured items, where trust was positive but more moderate than adoption, and where non-trivial minorities still reported score-oriented pressure or uncertainty about how to revise. These concerns meaningfully bound the claims, but they do not overturn the overall pattern that participants generally understood, used, and partly carried forward the feedback.

Overall, the questionnaire findings strengthen the interpretation of the artifact-based RQ1--RQ3 results. They suggest that the observed improvements were not merely score-level artifacts: participants generally reported understanding the feedback, translating it into concrete revisions, and reusing some of the learned practices in later tasks. The remaining concerns are therefore best interpreted as design-improvement opportunities for the feedback mechanism, rather than as evidence against its practical value.

\section{Discussion}
In this section, we synthesize the artifact-based and questionnaire-based evidence, discuss what the results imply for feedback integration in crowdsourced testing workflows, clarify how this study extends prior work, and outline the main limitations of the current design.

\subsection{Overall Interpretation of the Findings}
Taken together, the results support a consistent workflow-level interpretation. The clearest signal is not only that agent-generated feedback improves revised reports within the same round, but also that some of the reinforced practices carry over to later tasks. In other words, the assessment backbone did not merely assess completed reports after the fact; when its outputs were converted into actionable tester-facing feedback, it influenced how reports were subsequently produced.

This interpretation is supported by both the artifact-based and questionnaire-based evidence. The artifact-based analyses showed immediate revision gains (RQ1), improved first submissions on a new task after prior feedback exposure (RQ2), and positive longer-term transfer on a later application (RQ3). The questionnaire findings in Section~\ref{sec:questionnaire-findings} are consistent with this pattern: respondents generally reported understanding the feedback, translating it into concrete report revisions, and carrying some of the learned practices into later tasks. The overall picture is therefore one of \emph{feedback-enabled quality improvement} rather than score reporting alone.

At the same time, the mechanism should not be interpreted as uniformly or automatically effective. The results are better described as showing \emph{partial but meaningful transfer} in the studied setting. Improvement was broad, but not identical across metrics, tasks, and individuals. This heterogeneity is important, because it suggests that workflow-integrated agent feedback can be practically valuable even when it does not function as a frictionless or universally optimal intervention.

\subsection{Why the Feedback Helped, and Where It Fricted}
A plausible explanation for the positive results is that the feedback operated at two complementary levels. At the micro level, it reinforced report-writing practices such as clearer steps, more explicit expected results, and more complete defect reproduction information. At the macro level, it encouraged more deliberate checking of requirement coverage and more systematic attention to uncovered functional points. This two-level structure helps explain why gains were observed in both Textuality and Adequacy rather than in wording quality alone.

The questionnaire evidence further clarifies the mechanism. Participants did not simply report that the feedback was readable; many reported that they acted on it. In particular, the reported rates of revising wording/structure/details and adding missing test points or requirement coverage were both high, and the semi-open responses indicate that omitted requirement points, clearer steps, expected results, and more systematic coverage checking were among the most salient and reusable elements. This is consistent with the view that the intervention worked by making otherwise overlooked report-quality and coverage gaps visible in a form that testers could often translate into concrete actions.

However, the same evidence also helps explain why transfer was not perfectly uniform. First, some metrics exhibited ceiling effects: when baseline performance was already high, especially on adequacy, there was less room for visible improvement. Second, task characteristics mattered. In particular, the relatively familiar workflows of APP2 appear to have compressed between-group differences in adequacy, making transfer easier to observe in textuality than in coverage. Third, several respondents reported a translation problem rather than a rejection problem: they accepted the need for revision, but sometimes found it difficult to operationalize high-level coverage reminders into concrete test points. This is consistent with the concern items in the questionnaire, where adoption was stronger than trust and where some respondents still reported uncertainty about how to revise.

Accordingly, the feedback mechanism is best understood as useful but not frictionless. Its value does not depend on participants uniformly trusting every suggestion. Rather, its practical effectiveness appears to come from the fact that many participants could still convert the feedback into concrete report-level and coverage-level actions, even though some suggestions remained too generic, insufficiently grounded, or harder than desired to execute.

\subsection{Implications for Crowdsourced Testing Workflows}
The most immediate implication of these findings concerns crowdsourced testing platforms. Prior work on automated assessment for crowdsourced reports is naturally aligned with downstream review support: once reports are submitted, assessment can help developers or platform operators review, filter, rank, and prioritize them. The present study suggests that there is also value in moving the intervention upstream. Instead of using the assessment backbone only to sort reports after submission, platforms may also use it to improve report quality before those reports reach developers.

This distinction matters in practice. Downstream filtering can reduce review burden, but it does not necessarily improve the quality of what the crowd produces next. In contrast, tester-facing in-situ feedback may create a more constructive loop in which assessment serves not only as a selection mechanism but also as a means of improving future submissions. In the studied setting, this loop appears particularly promising for improving report completeness, testcase clarity, expected-result articulation, and requirement-coverage checking.

At the same time, the implications should be bounded to the workflow actually studied here. Our evidence most directly supports \emph{crowdsourced functional testing} on interactive applications with requirement documents and report artifacts of the kind used in this study. It does not yet justify broad claims about all software-engineering workflows, all artifact types, or all team settings. A more defensible implication is therefore that the current results motivate further study of feedback-oriented agent integration in other settings, rather than directly establishing general applicability beyond crowdsourced testing.

\subsection{Relation to Prior Work: More Than a Judge}
This study should be understood as extending the role established in prior work rather than re-validating the same assessment function. Prior work demonstrated that the assessment backbone can serve as a reliable and efficient assessment instrument for crowdsourced testing reports~\cite{ase2025multi}. That result was a necessary prerequisite: without sufficiently trustworthy assessment outputs, using those outputs as feedback in a human workflow would have little practical value.

The contribution of the present article is different. Here, the focus is not on whether the agents can score reports consistently, but on what happens when the same assessment foundation is embedded into the tester's workflow as actionable feedback. In this sense, the backbone becomes \emph{more than a judge}: it retains its evaluative role, but also becomes a feedback provider that can shape report production. The empirical contribution of this journal article therefore lies in showing that a previously validated assessment mechanism can also have meaningful workflow impact on human testers, including immediate revision gains, improved first submissions on later tasks, and evidence of partial skill transfer.

This shift from post-hoc assessment to in-situ feedback is also important conceptually. It changes the role of the agent from a downstream screening aid into a participant in upstream quality improvement. That does not mean the system should be romanticized as a full teammate or tutor in the broadest sense. Rather, the present evidence supports a more specific and defensible claim: in the studied setting, agent-generated feedback can function as a practical, workflow-integrated intervention that helps testers improve how they write and structure reports and how they check requirement coverage.

\subsection{Threats to Validity}
While the study provides encouraging evidence for workflow-integrated agent feedback, several limitations should be acknowledged.

\textbf{Internal validity.} A primary threat is the possibility of general learning effects across stages. Participants may improve partly because they gain experience by performing multiple related tasks, not only because of the feedback intervention itself. We mitigated this threat through the staged A/B design, which separates immediate revision effects, new-task first-submission contrasts, and later transfer comparisons. Even so, later-stage improvements should be interpreted as reflecting a combination of feedback-driven learning and broader task exposure rather than a perfectly isolated causal effect.

A second internal-validity issue is the use of external AI tools during the four-stage tasks. Questionnaire responses indicate that other AI use was common, although typically partial rather than pervasive. This means that external AI assistance cannot be treated as absent. At the same time, the self-reports suggest that it more often supported comprehension, planning, or wording than fully replacing report production. We therefore treat external AI use as a real but bounded confound rather than as a negligible factor.

\textbf{Construct validity.} Our primary outcome measures are derived from the assessment backbone itself, especially the Textuality and Adequacy dimensions. Prior work established the reliability of this backbone with respect to expert judgment~\cite{ase2025multi}, which is why it can serve as the measurement foundation of the present study. Nonetheless, the current design still evaluates improvement according to this particular operationalization of report quality. In other words, the study most directly shows improvement in report quality as defined by this framework, not across every possible dimension of real-world report usefulness.

A related construct-validity concern is that the same assessment foundation underlies both feedback generation and outcome measurement. This raises the possibility that some of the observed gains reflect increasing alignment with the framework's own quality criteria. The questionnaire findings partly mitigate this concern, because they suggest that participants experienced the feedback as useful in concrete writing and coverage-planning terms rather than purely as score optimization. Even so, this concern cannot be eliminated entirely and should remain part of the interpretation.

\textbf{External validity.} The generalizability of the findings is limited in several ways. First, the study involved 20 testers from one crowdsourced testing platform, and only 17 of those 20 completed the questionnaire. Although the respondent subset covered both study groups and was used only as complementary evidence, non-response bias remains possible, and the sample should not be treated as fully representative of broader crowdsourcing populations. Second, the participant pool is best characterized as a trained yet heterogeneous crowd-testing cohort rather than as professional software testers. The findings therefore transfer most naturally to similar crowd-testing settings, not automatically to industrial test teams or other developer populations.

Third, the study used only three applications, all of which are interactive applications with requirement artifacts suited to report-oriented functional testing. This is an appropriate scope for the present study, but it leaves open how the mechanism would behave for other artifact types or software domains, such as API-heavy systems, embedded software, safety-critical systems, games, or large-scale enterprise backends. Fourth, the questionnaire evidence is self-reported and therefore subject to recall error and social-desirability effects. The semi-open items also yielded limited free-text elaboration, so the qualitative analysis is necessarily lightweight and option-driven rather than interview-depth evidence.

Despite these limitations, the combination of staged artifact-based comparisons and complementary questionnaire evidence provides a reasonably strong basis for the central claim of this paper: in the studied crowdsourced testing setting, agent-generated in-situ feedback can improve report production and can leave at least partial effects that carry into later tasks.

\section{Conclusion}
This paper extends prior work on a multi-dimensional assessment backbone for crowdsourced testing reports into a different question and a different workflow role. Prior work established that the backbone can assess reports reliably and efficiently. Building on that foundation, the present study investigated whether those assessment outputs can also improve human work when they are delivered as actionable, tester-facing feedback in a realistic crowdsourced testing workflow.

Through a controlled four-stage human-subject study, we found consistent evidence of workflow-level value. Agent-generated feedback helped testers improve the quality of existing reports within the same round, was associated with better first submissions on later tasks, and showed evidence of partial but meaningful transfer to a third application. Questionnaire findings from the respondent subset reinforced this interpretation by suggesting that participants generally understood the feedback, acted on it in revision, and carried some of the reinforced practices into later tasks, while also revealing remaining friction in specificity, grounding, and execution support.

Taken together, these results suggest that assessment agents in crowdsourced testing can play a role that is more than post-hoc judging. In the studied setting, the same assessment foundation can also function as a workflow-integrated feedback mechanism that supports upstream report-quality improvement. More broadly, this study points toward a practical direction for agentic AI in software work: not only automating assessment after artifacts are produced, but also helping shape how those artifacts are produced in the first place.

Future work should test this design in larger and more diverse participant pools, extend it to other software domains and artifact types, and explore how feedback specificity, grounding, and adaptivity affect longer-term human learning in workflow-integrated agent settings.

\section{DATA AVAILABILITY}
Our experimental materials are available at https://github.com/LLMJudge/MoreThanAJudge.

\begin{acks}
The authors would like to thank all anonymous reviewers for their insightful comments. This work was partially supported by the National Key R\&D Program of China under Grant 2024YFF0908000.
\end{acks}

\bibliographystyle{ACM-Reference-Format}
\bibliography{sample-base}

\end{document}